\documentclass[a4paper, 11pt]{article} 
\usepackage{a4wide}
\usepackage{longtable}
\usepackage{graphicx}
\usepackage{bm}
\usepackage{titlesec}

\usepackage[shortlabels]{enumitem}

\usepackage[hyphens]{url}
\usepackage{appendix}
\usepackage{enumitem}

\usepackage[nottoc]{tocbibind}

\usepackage[margin=1in]{geometry}
\usepackage{setspace}
\usepackage[ruled, vlined]{algorithm2e}
\usepackage{multirow}
\usepackage{amsmath, amssymb, amsthm, float}
\usepackage{mathtools}
\usepackage{caption}
\usepackage{color}

\usepackage{hyperref}
\usepackage[]{algorithm2e}
\usepackage{booktabs}

\allowdisplaybreaks
\usepackage{mathtools}

\newcommand{\tcr}[1]{\textcolor{black}{#1}}

\theoremstyle{definition}
\newtheorem{theorem}{Theorem}

\newtheorem{example}{Example}
\newtheorem{assumption}{Assumption}
\theoremstyle{remark}
\newtheorem{remark}{Remark}

\usepackage{amsfonts}
\usepackage{dsfont}
\usepackage{natbib}

\def\wh{\widehat}
\def\wt{\widetilde}

\newcommand{\mc}{\mathcal}
\newcommand{\E}[0]{\mathbb{E}}
\newcommand{\Var}[0]{\mathsf{Var}}

\newcommand{\p}{\mathsf{P}}
\newcommand{\vep}{\varepsilon}
\renewcommand{\l}{\left}
\renewcommand{\r}{\right}
\newcommand{\Z}{\mathbb{Z}}
\DeclareMathOperator*{\argmax}{arg\,max}

\title{Robust multiscale estimation of time-average variance for time series segmentation}

\author{Euan T. McGonigle$^1$ \and Haeran Cho$^2$}

\begin{document}

\maketitle

\footnotetext[1]{Institute for Statistical Science, School of Mathematics, University of Bristol.
Email: \url{euan.mcgonigle@bristol.ac.uk}.}

\footnotetext[2]{Institute for Statistical Science, School of Mathematics, University of Bristol.
Email: \url{haeran.cho@bristol.ac.uk}. Supported by Leverhulme Trust Research Project Grant RPG-2019-390.} 

\begin{abstract}
There exist several methods developed for the canonical change point problem of detecting multiple mean shifts, which search for changes over sections of the data at multiple scales. In such methods, estimation of the noise level is often required in order to distinguish genuine changes from random fluctuations due to the noise. When serial dependence is present, using a single estimator of the noise level may not be appropriate. Instead, \tcr{it is proposed} to adopt a scale-dependent time-average variance constant that depends on the length of the data section in consideration, to gauge the level of the noise therein. Accordingly, an estimator that is robust to the presence of multiple mean shifts is \tcr{developed}. The consistency of the proposed estimator \tcr{is shown} under general assumptions permitting heavy-tailedness, and its use with two widely adopted data segmentation algorithms, the moving sum and the wild binary segmentation procedures, \tcr{is discussed}.
The performance of the proposed estimator \tcr{is illustrated} through extensive simulation studies and on applications to the house price index and air quality data sets. 
\end{abstract}

\noindent%

\section{Introduction}

Dating back to the 1950s \citep{page1954continuous},
change point analysis has a long tradition in statistics. 
The area continues to be an active field of research due to its importance in many applications where data is routinely collected in highly nonstationary environments. 
%examples of which include genetics \citep{hocking2013learning} and climatology \citep{reeves2007review}.
Data segmentation, a.k.a.\ multiple change point detection, enables partitioning of a time series into stationary regions and thus provides a simple framework for modelling nonstationary time series. 

We consider the problem of detecting multiple change points in the piecewise constant mean of an otherwise stationary time series.
We briefly review the existing literature on multiple change point detection 
in the presence of serial dependence, and refer to \cite{aue2013} and \cite{truong2020selective} for a comprehensive overview.
One line of research takes a parametric approach
and simultaneously estimates the serial dependence and change points. For example, \cite{chakar2017robust}, \cite{fang2020detection} and \cite{romano2021detecting} assume an autoregressive (AR) model of order one, while \cite{lu2010mdl}, \cite{cho2021multiple} and \cite{gallagher2022autocovariance} permit an AR model of arbitrary order.

Another line of research focuses on extending the use of the methodologies
developed for independent data to time series settings.
\cite{lavielle2000} and \cite{cho2021data} adopt information criteria originally developed for a sequence of independent, Gaussian random variables \citep{yao1988}, to data exhibiting serial correlations and heavy tails,
which requires the choice of an appropriate penalty that depends on the tail behaviour of noise. 
\cite{tecuapetla2017autocovariance}, \cite{eichinger2018mosum}, \cite{dette2020multiscale} and \cite{chan2022optimal}
propose estimators of the long-run variance (LRV) for quantifying the level of noise
that are robust to the presence of multiple mean shifts. 
% Such estimators are then combined with a multiple change point detection algorithm.
We also note that \cite{wu2019mace} and \cite{zhao2021segmenting} extend self-normalisation-based change point tests % \citep{shao2010testing, pevsta2020nuisance}
to the data segmentation problem.

In this paper, we propose a robust estimator of the
scale-dependent time-average variance constant (TAVC, \citeauthor{wu2009recursive}, \citeyear{wu2009recursive}).
It is closely related to the literature on estimation of the LRV,
namely $\sigma^2 = \lim_{N \to \infty} \Var(N^{-1/2} \sum_{t = 1}^N \vep_t)$ for a stationary time series $\{\vep_t\}_{t \in \Z}$,
but distinct in that our interest lies in estimating 
\begin{align}
\label{eq:tavc}
\sigma^2_L = \Var\l( \frac{1}{\sqrt L} \sum_{t = 1}^L \vep_t \r),
\end{align}
for a given scale $L$.
We argue that such a scale-dependent TAVC estimator is well-suited 
to be combined with a class of {\it multiscale} change point detection methodologies, examples of which include
the moving sum (MOSUM) procedure \citep{eichinger2018mosum} and the wild binary segmentation (WBS, \citeauthor{fryzlewicz2014wild}, \citeyear{fryzlewicz2014wild}) algorithm. 
Such approaches {\it locally} apply change point tests for single change point detection, to data sections of varying lengths and achieve good adaptivity in multiple change point detection \citep{cho2021data}.
We motivate the use of scale-dependent TAVC in~\eqref{eq:tavc} in combination with such multiscale methods in the following examples.

\begin{figure}[H]
\centering
\includegraphics[width = 0.8\textwidth]{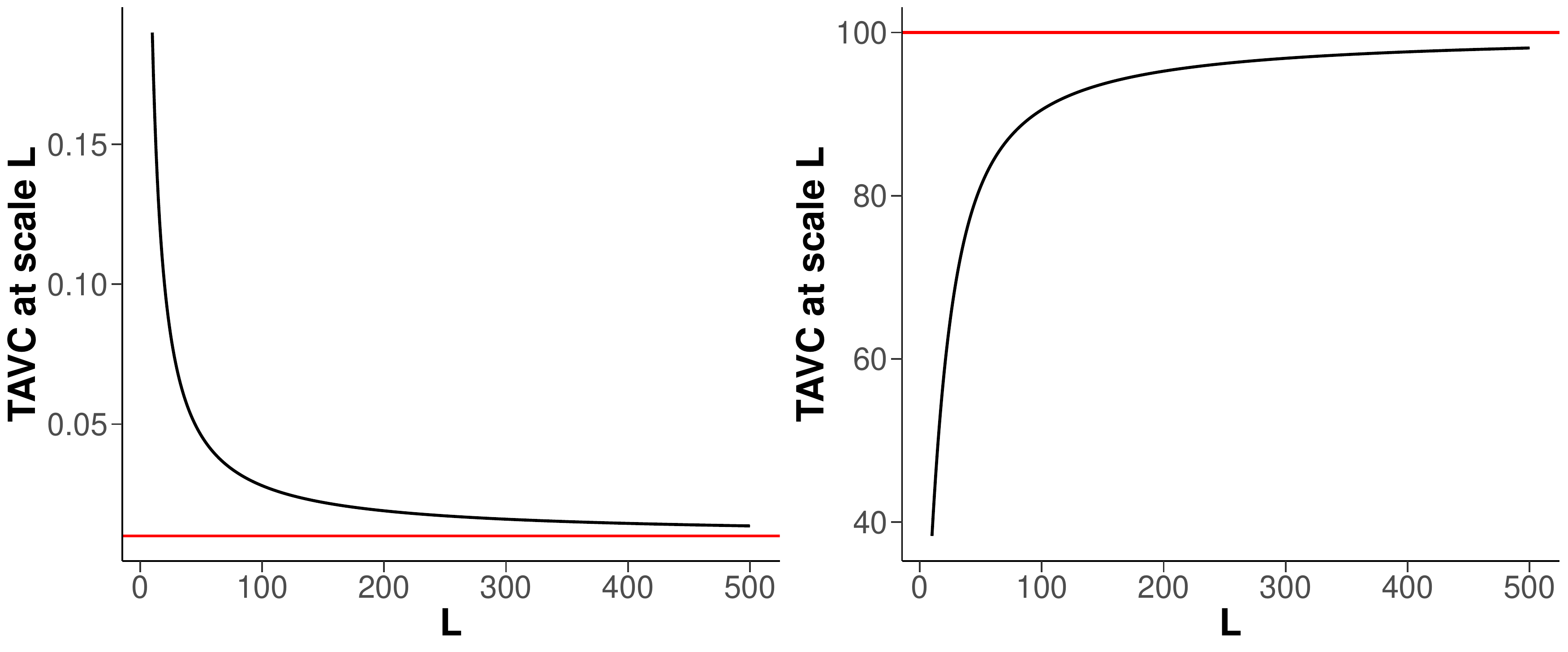}
\caption{\tcr{TAVC $\sigma^2_L$ for increasing $L$ (black line) computed on the MA(1) process $\varepsilon^{(1)}_t = W_t - 0.9 W_{t-1}$ (left) and the AR(1) process $\varepsilon^{(2)}_t = 0.9 \varepsilon^{(2)}_{t-1} + W_{t}$ (right).
The respective LRV is given by the horizontal red line.}}
\label{fig:tavc-lrv}
\end{figure}

\begin{figure}[H]
\centering
\includegraphics[width = 0.9\textwidth]{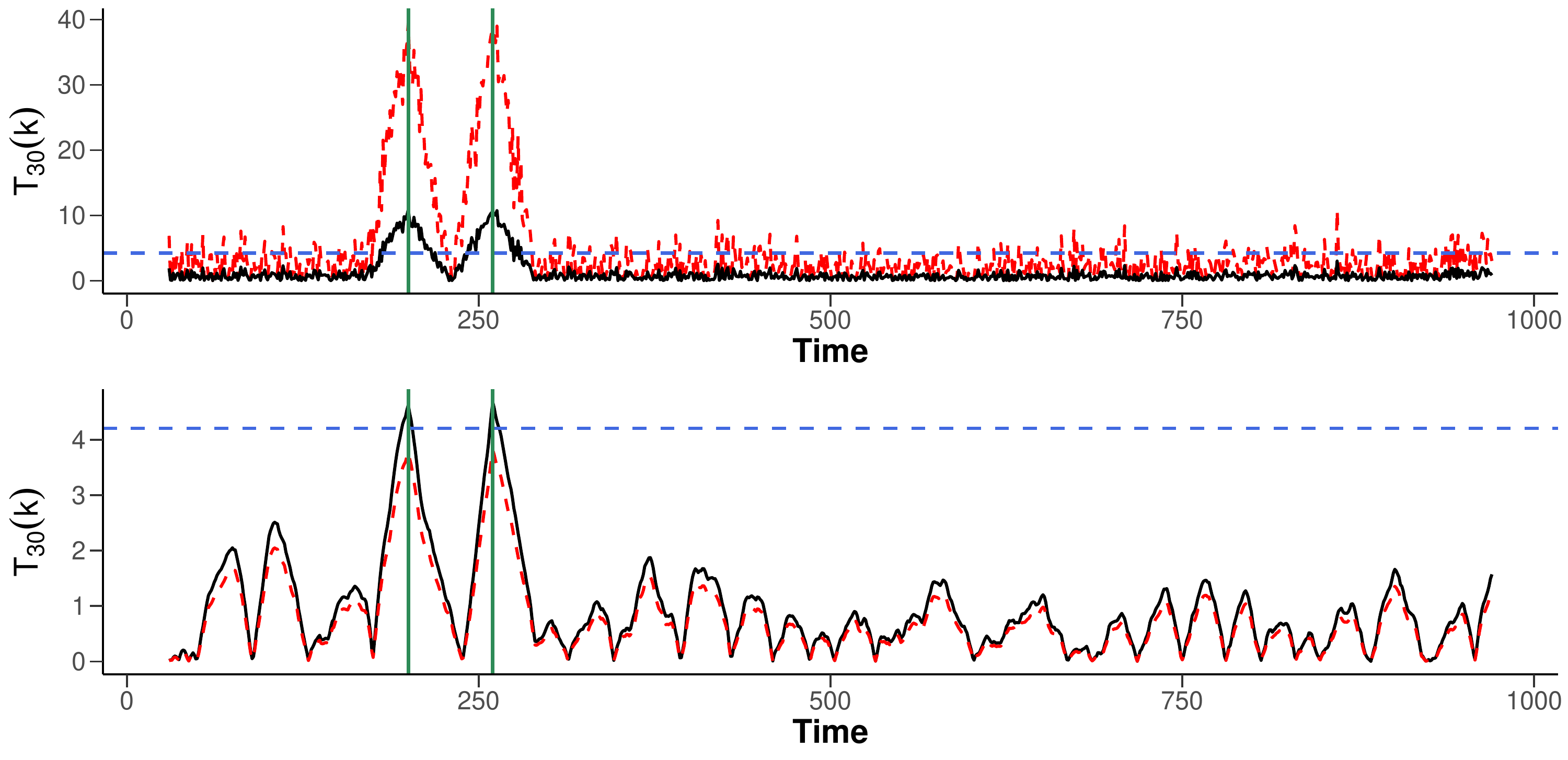}
\caption{\tcr{Scaled MOSUM detector statistics computed on the data generated by adding $\{\varepsilon^{(1)}_t\}_{t = 1}^n$ (top) and $\{\varepsilon^{(2)}_t\}_{t = 1}^n$ (bottom), to a piecewise constant signal with two change points ($n = 1000$, green vertical lines denote the true change locations). 
The detector statistics scaled by the proposed TAVC estimator (black solid line) and true LRV (red dashed line) are plotted. Dashed blue horizontal line denotes a theoretically motivated threshold at $5$\% significance level.}}
\label{fig:example-stats}
\end{figure}

\begin{example} \label{ex:one}
Consider an MA($1$) process $\varepsilon^{(1)}_t = W_t - 0.9 W_{t-1}$
and an AR($1$) process $\varepsilon^{(2)}_t = 0.9 \varepsilon^{(2)}_{t-1} + W_{t}$,
where $\{W_t\}_{t \in \Z}$ is a white noise process with $\Var(W_t) = 1$.
\tcr{Figure~\ref{fig:tavc-lrv} shows the TAVC of $\{\varepsilon^{(1)}_t\}_{t \in \Z}$ and $\{\varepsilon^{(2)}_t\}_{t \in \Z}$ for increasing $L$, along with the true LRV, which highlights the large gap between $\sigma^2_L$ and $\sigma^2$ particularly at a small scale $L$. 
This discrepancy has an impact on the performance of change point detection methods.}

\tcr{Figure~\ref{fig:example-stats} further illustrates this point by plotting the MOSUM detector statistics generated with a moving window of length $30$ (see Equation~\eqref{eq:mosum} for its definition), on the data generated by adding $\{\varepsilon^{(1)}_t\}_{t = 1}^n$ and $\{\varepsilon^{(2)}_t\}_{t = 1}^n$ to a piecewise signal with two change points at times $\tau_1 = 200$ and $\tau_2 = 260$, respectively ($n = 1000$).
Then, the detector statistics are scaled by the proposed estimator of TAVC (solid) and the true LRV (dashed).
For accurate detection of the change points, we expect that a sequence of appropriately scaled detector statistics forms two prominent peaks at the change points that exceed a theoretically motivated threshold, while away from the change points, detector statistics remain below the threshold, see Section~\ref{sec:mosum} for further details.
In combination with the TAVC estimator, the detector statistics exhibit the desired behaviour such that both change points are detectable from the scaled MOSUM statistics.
However, due to the lack of adaptivity of the LRV to the scale-dependent variability of the detector statistics, its use leads to either a large number of false positives (spurious peaks above threshold, see the top panel of Figure~\ref{fig:example-stats}), or false negatives (MOSUM detector statistics scaled by the LRV do not exceed the threshold near $\tau_i, \, i = 1, 2$, see the bottom panel of Figure~\ref{fig:example-stats}).}
\end{example}

Example~\ref{ex:one} demonstrates that adopting the global LRV may fail to reflect the degree of variability in the local data sections that are used in computing change point detector statistics adopted by multiscale data segmentation algorithms,
which in turn may result in false negatives or positives. 
% (failure in detecting genuine changes) or false positives (detection of spurious estimators).
Moreover, when the LRV is close to zero as in the case of $\{\vep^{(1)}_t\}_{t \in \Z}$ in Example~\ref{ex:one}, some estimators of the LRV have been observed to take negative values \citep{huvskova2010note}, which further makes their use in change point problems undesirable.

To ensure that the scale-dependent TAVC estimator is robust to the mean shifts, we adopt the robust $M$-estimation framework of \cite{catoni2012challenging}, which was first proposed in the independent setting for mean and variance estimation and further extended to the serially dependent setting for LRV estimation in \cite{chen2021inference}. 
We establish the consistency of the proposed robust estimator of scale-dependent TAVC under general conditions permitting heavy tails and serial dependence decaying at a polynomial rate.
Then, we discuss its application with multiscale change point detection methods such as the MOSUM procedure and the WBS algorithm,
and provide a heuristic approach to accommodate local stationarity in the data.

The remainder of the article is organised as follows. 
Section~\ref{sec:tavc} introduces the scale-dependent TAVC and
its robust estimator and establishes its consistency. 
Section~\ref{sec:multiscale} discusses its application with multiscale data segmentation algorithms and an extension to local stationarity.
In Section~\ref{sec:num}, we examine the performance of the proposed methodology
on simulated data sets and two real data examples on house price index and air quality.
Section~\ref{sec:conc} concludes the paper.
All proofs, algorithmic descriptions of multiscale change point methods
and additional numerical results are given in the appendix. 
Accompanying \verb!R! software implementing the methodology is available from \url{https://github.com/EuanMcGonigle/TAVC.seg}.

\section{Scale-dependent TAVC and its robust estimation}
\label{sec:tavc}

%In this section we describe the change in mean model and its assumptions and introduce the robust multiscale estimator of the TAVC that is used to standardise the test statistics.

\subsection{Multiscale change point detection in the mean}

We consider the problem of multiple change point detection
under the following model:
\begin{align}
\label{eq:model}
X_{t} = f_{t} + \varepsilon_{t} = f_{0} + \sum_{i=1}^{q} \mu_{i} \cdot \mathbb{I}(t \geq \tau_{i}+1) + \varepsilon_{t}, \quad t = 1, \ldots, n.
\end{align}
Under the model~\eqref{eq:model}, the piecewise constant signal $f_{t}$ contains $q$ change points at locations $\tau_{i}$, $i = 1, \ldots, q$, with the notational convention that $\tau_{0}=0$ and $\tau_{q+1} = n$. 
The errors $\{ \varepsilon_{t} \}_{t=1}^{n}$ are assumed to be a (weakly) stationary time series satisfying $\mathbb{E} (\varepsilon_{t}) = 0$
with finite LRV $\sigma^2 = \lim_{N \to \infty} \Var(N^{-1/2} \sum_{t = 1}^N \vep_t) \in (0, \infty)$,
and are permitted to be serially correlated and heavy-tailed
(see Assumption~\ref{assum:noise} below). 
Our aim is to consistently estimate the total number and the locations of the change points.
\tcr{While our primary focus is on detecting changes in the mean, it does not exclude the possibility of applying the proposed method to detecting changes in stochastic properties other than the mean via suitable data transformation as outlined in \cite{cho2021data}.}

A common approach to this problem is closely related to the change point testing literature,
which scans the data for the detection and estimation of multiple change points
by {\it locally} applying a test well-suited for detecting a single change.
Such a procedure typically involves comparing a test statistic of the form
$\wh{\sigma}_{s, e}^{-1} \vert \mathcal{T}_{s, k, e} \vert$ to a threshold, say $D$.
Here,
\begin{align}
\label{eq:cusum-stat}
\mathcal{T}_{s, k, e} = \sqrt{ \frac{(k-s)(e-k)}{e-s}} \left(  \frac{1}{k-s} \sum_{t=s+1}^{k} X_{t}  -  \frac{1}{e-k} \sum_{t=k+1}^{e} X_{t} \right)
\end{align}
denotes a change point detector evaluated at some locations
$0 \le s < k < e \le n$, which are determined in a method-specific way
(see Section~\ref{sec:multiscale} below),
$\sigma_{s, e}^2$ denotes a measure of variability
in the data section $\{X_t\}_{t = s + 1}^e$, and
$\wh\sigma_{s, e}^2$ its estimator.
Under the stationarity assumption on $\{\vep_t\}_{t = 1}^n$,
a natural choice is $\sigma^2_{s, e} = \sigma^2_{e - s}$,
the scale-dependent TAVC defined in~\eqref{eq:tavc} with $L = e - s$ as the scale.
Then, if $\{X_t\}_{t = s + 1}^e$ 
does not contain any change point well within the interval, we expect
$\wh{\sigma}_{s, e}^{-1} \vert \mathcal{T}_{s, k, e} \vert \le D$,
while it signals the presence of such a change point
when $\wh{\sigma}_{s, e}^{-1} \vert \mathcal{T}_{s, k, e} \vert > D$.
The key challenge lies in separating genuine mean shifts from the natural fluctuations due to serial correlations,
which requires a careful selection of the estimator $\wh\sigma_{s, e}^2$
that correctly captures the variability in the section of the data under consideration.

It is well-documented that multiscale application of such a test
on data sections of varying lengths,
improves the adaptivity of the change point methodology to detect both large, frequent changes 
and small changes over long stretches of stationarity \citep{cho2021data}.
For such a multiscale procedure, 
Example~\ref{ex:one} demonstrates the potential pitfalls associated with 
using an estimator of the global LRV in place of $\wh\sigma_{s, e}^2$,
regardless of the length of the interval on which the detector statistic in~\eqref{eq:cusum-stat} is computed.
In the next section, we propose an estimator of
the scale-dependent TAVC $\sigma^2_L$ in~\eqref{eq:tavc}
that is robust to the presence of multiple mean shifts,
for the standardisation of multiscale change point detectors.

\subsection{Robust estimation of multiscale TAVC}
\label{sec:robust}

For notational convenience, suppose that $L$ is an even number, and let $G = L/2$ denote the block size. Then, for some starting point $b \in \{0, \ldots, G - 1\}$ with number of blocks $N_1(b) = \lfloor (n-b-G)/G \rfloor$, we define
\begin{align*}
\bar{X}_{j, b} = \frac{1}{G} \sum_{t = jG + b + 1}^{(j+1)G + b} X_t,
\quad \text{and} \quad
\xi_{j, b} =  \frac{G(\bar{X}_{j, b} - \bar{X}_{j-1, b})^2}{2}
\text{ for } j = 1, \ldots, N_1(b).
\end{align*}
Analogously, we define
\begin{align*}
\bar{\vep}_{j, b} = \frac{1}{G} \sum_{t=jG + b + 1}^{(j+1)G + b} \vep_t
\quad \text{and} \quad
\wt\xi_{j, b} =  \frac{G(\bar{\vep}_{j, b} - \bar{\vep}_{j-1, b})^2}{2}.
\end{align*}
Then, the following sum
\begin{align}\label{no-change-est}
\wh{\wt\sigma}^{2}_L = \frac{1}{N_1(b)} \sum_{j=1}^{N_1(b)} \wt\xi_{j, b},
\end{align}
takes into account the temporal dependence in the local data sections of length $L = 2G$.
Further, we have
$\E(\wh{\wt\sigma}^2_L) - \sigma^2_L = o(1)$ for $L \to \infty$
(see Theorem~\ref{thm:one} below),
such that $\wh{\wt\sigma}^2_L$ is indicative of the level of variability $\sigma^2_L$ albeit being inaccessible (as it is defined with $\bar{\vep}_{j, b}$ in place of $\bar{X}_{j, b}$).
Its accessible counterpart, $N_1(b)^{-1} \sum_{j = 1}^{N_1(b)} \xi_{j, b}$, on the other hand, is typically biased due to the mean shifts
and thus is inappropriate as an estimator of the scale-dependent TAVC.

To obtain an estimator that is robust to multiple mean shifts,
we adopt the robust $M$-estimation framework of \cite{catoni2012challenging}.
Let $\phi$ denote a non-decreasing influence function as
\begin{align}
\label{eq:envelope}
\phi(x) = \begin{cases}
- \log(2) & \text{for } x \leq -1 , \\
\log(1 + x + x^{2}/2 ) & \text{for } -1 \leq x \leq 0, \\
- \log( 1 - x + x^{2}/2 ) & \text{for } 0 \leq x \leq 1, \\
\log(2) & \text{for } x \geq 1.
\end{cases}
\end{align}
The robust estimator of the TAVC at scale $L$ and starting point $b$, denoted $\wh{\sigma}^2_{L, b}$, is defined as the solution of the $M$-estimation equation
\begin{align}\label{eq:m:est}
h_{L, b} (u) = \frac{1}{N_1(b)} \sum_{j = 1}^{N_1(b)} \phi_v\l( \xi_{j, b} - u \r) = 0,
\end{align}
where $\phi_{v}(x) = v^{-1}\phi(vx)$ for some $v > 0$;
we specify the condition on $v$ later.
If there are multiple solutions to Equation~\eqref{eq:m:est}, any of them may be chosen. 

\subsection{Theoretical properties}

We establish the consistency of the scale-dependent TAVC estimator 
under the following assumption on the error process $\{ \varepsilon_t \}_{t=1}^n$.
\begin{assumption}
\label{assum:noise}
\hfill
\begin{enumerate}[label = (\roman*)]
\item \label{assum:noise:one} 
We assume that $\varepsilon_t = \sum_{k=0}^\infty {a}_k {\eta}_{t-k}$, 
where $\{\eta_t\}_{t \in \Z}$ is a sequence of i.i.d.\ random variables
and $\vert {a}_k \vert \leq \Xi (k + 1)^{-\beta}$ for some constants $\beta > 2.5$ and $\Xi > 0$ for all $k \ge 0$.
\item \label{assum:noise:two} 
There exists a fixed constant $c_\sigma > 0$ such that
$\sigma^2 = ( \sum_{k\geq 0} a_k )^2$ satisfies $\sigma^2 \ge c_{\sigma}$.
\item \label{assum:noise:three} We operate under {\it either} of the following two conditions on the distribution of $\{\eta_t\}_{t \in \Z}$.
\begin{enumerate}[label = (\alph*)]
\item \label{cond:moment} There exists a fixed constant $r > 4$ such that
$\Vert \eta_{1} \Vert_r = ( \mathbb{E} ( |\eta_{1} |^r) )^{1/r} <\infty$.
\item \label{cond:subexp} There exist fixed constants $C_\eta > 0$ and $\kappa \ge 0$ such that
$\Vert \eta_1 \Vert_r \le C_\eta r^\kappa$ for all $r \ge 1$. 
\end{enumerate}
\end{enumerate}
\end{assumption}

\tcr{The linearity of the process $\{\vep_t\}_{t = 1}^n$ assumed in Assumption~\ref{assum:noise}~\ref{assum:noise:one} facilitates the controlling of the functional dependence in $\{\xi_{j, b}\}_{j = 1}^{N_1(b)}$.
The condition} permits the temporal dependence to decay at an algebraic rate. Assumption~\ref{assum:noise}~\ref{assum:noise:two} is made to ensure that the LRV is well-defined. 
Assumption~\ref{assum:noise}~\ref{assum:noise:three}~\ref{cond:moment} allows heavy-tailed $\{\vep_t\}_{t = 1}^n$, 
while~\ref{cond:subexp} assumes a stronger condition
that requires sub-Weibull \citep{wong2020lasso} tail behaviour on $\{\eta_t\}_{t \in \Z}$ which includes sub-Gaussian ($\kappa = 1/2$) and sub-exponential ($\kappa = 1$)
distributions as special cases.

\tcr{For sequences of positive numbers $\{a_n\}$ and $\{b_n\}$, write $a_n \asymp b_n$ if there exists some positive constants $C_1$ and $C_2$ such that $C_1 \leq a_n/b_n \leq C_2$ as $n \to \infty$}. The following theorem establishes the consistency of the estimator of the scale-dependent TAVC, 
see Appendix~\ref{sec:proof} for the proof.

\begin{theorem}\label{thm:one}
Suppose that Assumption~\ref{assum:noise} holds, and define 
\begin{align*}
\wt{\sigma}^2_L = \Var \left( \frac{1}{\sqrt{L}} \left( \sum_{t=1}^{G} \varepsilon_t - \sum_{t=G+1}^L \varepsilon_t \right) \right) 
\end{align*}
for $L = 2G$.
Then, provided that $v \asymp \sqrt{q G/n}$,
the estimator $\wh{\sigma}^{2}_{L, b}$ satisfies
\begin{align}
\left| \wh{\sigma}^{2}_{L, b} - \wt{\sigma}^{2}_L \right| &= \mathcal{O}_P\l( \sqrt{ \frac{Lq}{n} } + \max\l\{ \l( \frac{L}{n} \r)^{\frac{r-2}{r+2}}, \sqrt{\frac{L \log(n)}{n}} \r\} \r) \label{eq:thm:one} 
\end{align}
under Assumption~\ref{assum:noise}~\ref{assum:noise:three}~\ref{cond:moment}, and
\begin{align}
\left| \wh{\sigma}^{2}_{L, b} - \wt{\sigma}^{2}_L \right| &= \mathcal{O}_{P} \left( \sqrt{ \frac{Lq}{n} } +  {\sqrt{ \frac{L \log^{4\kappa + 3} (n)}{n} }}\right) 
\label{eq:thm:two}
\end{align}
under Assumption~\ref{assum:noise}~\ref{assum:noise:three}~\ref{cond:subexp}, for any fixed $b \in \{0, \ldots, G - 1\}$.
In addition, $\wt{\sigma}^2_L$ satisfies
\begin{align}
\left| \wt{\sigma}^{2}_{L} - {\sigma}_L^{2} \right| = \mathcal{O}(L^{-1}).
\label{eq:thm:three} 
\end{align}
\end{theorem}

Theorem~\ref{thm:one} shows that the proposed robust estimator consistently estimates the TAVC at scale $L$.
The estimation error is decomposed into the error from approximating $\sigma_L^2$ by $\wt\sigma_L^2$ in~\eqref{eq:thm:three}, and
that in estimating $\wt{\sigma}_L^2$ by $\wh{\sigma}_{L,b}^2$.
In deriving the second error, we make explicit the influence of multiple mean shifts on the estimator by the term $(Lq/n)^{1/2}$ in~\eqref{eq:thm:one}--\eqref{eq:thm:two}, as well as the effect of the innovation distribution in the remaining terms.
A careful examination of the proof of Theorem~\ref{thm:one} shows that
\begin{align}
\left| \wt{\sigma}^{2}_{L} - {\sigma}^{2} \right| &= \mathcal{O}(L^{-1}),
\label{sigma-approx2}  
\end{align}
therefore as $L$ increases, the scale-$L$ TAVC approximates the global LRV as expected.

\begin{remark}[Maximum time-scale for TAVC estimation]
\label{rem:max}
The error due to approximating $\sigma_L^2$ with $\wt{\sigma}_L^2$ 
decreases with $L$ as in~\eqref{eq:thm:three}.
On the other hand, the error of estimating $\wt{\sigma}_L^2$ with $\wh{\sigma}_{L, b}^2$ increases with $L$ as in~\eqref{eq:thm:one}--\eqref{eq:thm:two};
this is attributed to the effect of mean shifts that grows with $L$,
and the decrease in the number of available blocks.
To balance between the two,
we suggest setting a maximum time-scale, say $M$, to be used in combination with a multiscale change point detection algorithm.
That is, when the change point detector $\mc T_{s, k, e}$ involves 
$e - s \le M$, we scale the detector with the estimator of $\sigma^2_{s - e}$, the TAVC at the corresponding scale $L = e - s$.
On the other hand, if $e - s > M$, we propose to scale the detector
with the estimator of $\sigma^2_{M}$, the TAVC at the maximum time-scale $M$,
which satisfies $\vert \sigma^2_{e - s} - \sigma^2_M \vert = \mathcal{O} (M^{-1})$.
\end{remark}

\section{Applications and extensions}
\label{sec:multiscale}

We now describe explicitly how the robust estimator of the scale-dependent TAVC proposed in Section~\ref{sec:tavc}, is applied within the algorithms 
that scan moving sum (MOSUM) and cumulative sum (CUSUM) statistics of the form~\eqref{eq:cusum-stat}, for multiple change point detection. 

\subsection{MOSUM procedure}
\label{sec:mosum}

The MOSUM procedure \citep{chu1995mosum,eichinger2018mosum} 
evaluates the change point detector $\mc T_{s, k, e}$ over a moving window. 
For a given bandwidth $G$, the MOSUM detector for a change in mean at time point $k$ is given by
\begin{align}
\label{eq:mosum}
\mc T_{G}(k)  = \mathcal{T}_{k-G, k, k+G} =   \sqrt{\frac{G}{2}} \left(  \frac{1}{G} \sum_{t=k+1}^{k+G} X_{t} - \frac{1}{G} \sum_{t=k-G+1}^{k} X_{t}   \right), \quad G  \le k \le n - G.
\end{align}
\cite{eichinger2018mosum} propose to estimate the total number and the locations of multiple change points by
identifying all significant local maximisers of $\mc T_G(k)$, say $\wh k$, satisfying
\begin{align}
\label{eq:mosum:est}
\vert \mc T_G(\wh k) \vert > \wh\sigma \cdot D_n(G; \alpha) \quad \text{and} \quad
\wh k = {\arg\max}_{k: \, \vert k - \wh k \vert \le \eta G} \vert \mc T_G(k) \vert
\end{align}
for some $\eta \in (0, 1)$.
Here, $D_n(G; \alpha)$ denotes a critical value at a significance level $\alpha \in (0, 1)$,
which is drawn from the asymptotic null distribution of the MOSUM test statistic $\max_{G \le k \le n - G} \sigma^{-1} \vert T_G(k) \vert$ obtained
under mild conditions permitting heavy-tailedness and serial dependence, and
takes the form 
\begin{align*}
D_{n} (G, \alpha) = \frac{b_{G,n} + c_{\alpha}}{a_{G,n}} \quad \text{with} \quad 
c_{\alpha} = - \log \log \left( \frac{1}{\sqrt{1-\alpha}} \right),
\end{align*}
where $a_{G,n}$ and $b_{G,n}$ are known constants depending on $G$ and $n$ only. The single-bandwidth MOSUM procedure achieves consistency in estimating the total number and the locations of multiple change points,
provided that $\min_{1 \le i \le q} \mu_i^2 G \to \infty$ as $n \to \infty$ sufficiently fast while $2G \le \min_{0 \le i \le q} (\tau_{i + 1} - \tau_i)$, see Theorem~3.2 of \cite{eichinger2018mosum} and Corollary D.2 of \cite{cho2022two} for explicit conditions.
The requirement on $G$ indicates that the single-scale MOSUM procedure performs best with the bandwidth chosen as large as possible while avoiding situations where there are more than one change point within the moving window at any time. Consequently, it lacks adaptivity when the data sequence contains both large changes over short intervals and small changes over long intervals.

Multiscale extensions of the single-bandwidth MOSUM procedure,
i.e.\ applying the MOSUM procedure with a range of bandwidths 
and then combining the results, alleviate the difficulties involved in bandwidth selection and provide adaptivity.
In this paper, we consider the multiscale MOSUM procedure
combined with the `bottom-up' merging as proposed by \cite{messer2014multiple} (see also \cite{meier2021mosum}).
Denoting the range of bandwidths by $\mc G = \{G_h, \, 1 \le h \le H: \,
G_1 < \ldots < G_H\}$, 
let $\mc C(G)$ denote the set of estimators detected with some bandwidth~$G \in \mc G$.
Then, we accept all estimators in $\mc C(G_1)$ returned with the finest bandwidth $G_1$ to the set of final estimators $\mc C$ 
and, sequentially for $h = 2, \ldots, H$, accept $\wh k \in \mc C(G_h)$ if and only if
$\min_{k \in \mc C} \vert \wh k - k \vert > \eta G_h$
(with $\eta$ identical to that in~\eqref{eq:mosum:est}). That is, we only accept the estimators that do not detect the change points which have previously been detected at a finer scale.

We propose to apply the multiscale MOSUM procedure with bottom-up merging,
in combination with the robust estimator of multiscale TAVC as follows.
For each $G_h \in \mc G$, the TAVC at scale $L = 2 G_h$ is estimated
by $\wh{\sigma}^2_{2 G_h}$ solving~\eqref{eq:m:est}, provided that $2G_h \le M$.
Here, $M$ denotes the maximum scale which is set in relation to the sample size $n$, see Remark~\ref{rem:max}.
Then, we use $\wh{\sigma}_{2 G_h}$ in place of the global estimator $\wh\sigma$
in~\eqref{eq:mosum:est} to standardise the MOSUM detector $\mc T_{G_h}(k)$.
When $2G_h > M$, we use $\wh{\sigma}_M$ in place of $\wh\sigma_{2G_h}$
for MOSUM detector standardisation.
In doing so, we ensure that change point detectors at multiple scales are standardised
by the scale-dependent TAVC that accurately reflects the degree of variability over the moving window \tcr{(see Example \ref{ex:one})} , while taking into account the presence of possibly multiple mean shifts therein.
We refer to Algorithm~\ref{alg:mosum} in Appendix~\ref{appendix:alg} for the pseudocode of the multiscale MOSUM procedure with the robust estimator of scale-dependent TAVC.

\subsection{Wild binary segmentation}
\label{sec:wbs}

The binary segmentation algorithm \citep{scott1974cluster, vostrikova1981} and its extensions, such as wild binary segmentation (WBS, \citeauthor{fryzlewicz2014wild}, \citeyear{fryzlewicz2014wild}; \citeyear{fryzlewicz2020detecting}) and seeded binary segmentation \citep{kovcs2020seeded}, recursively search for multiple change points using the CUSUM statistic of the form~\eqref{eq:cusum-stat}, with $s$ and $e$ that are identified iteratively.
These methods have primarily been analysed for the data segmentation problem under~\eqref{eq:model}
assuming i.i.d.\ Gaussianity on the $\{\vep_t\}_{t = 1}^n$. 
Consequently, some robust estimators of $\Var(\vep_1)$ have been considered for standardising the CUSUM statistic.
Here, we discuss the application of the WBS2 algorithm \citep{fryzlewicz2020detecting} in the time series setting with the proposed robust estimator of the scale-dependent TAVC.

Let $\mc A_{s, e} = \{(\ell, r) \in \Z^2:\, s \le \ell < r \le e, \ r - \ell > 1\}$ 
denote the collection of all intervals within $\{s + 1, \ldots, e\}$ for some $0 \le s < e \le n$, and $\mc R_{s, e}$ denote its subset selected either randomly or deterministically
(see \cite{cho2021multiple} for one approach to deterministic grid selection)
with $\vert \mc R_{s, e} \vert = \min(R, \vert \mc A_{s, e} \vert)$ for some given $R \le n(n - 1)/2$. 
Starting with $(s, e) = (0, n)$, we identify
\begin{align}
\label{eq:wbs:est}
(s_\circ, k_\circ, e_\circ) = {\argmax}_{\substack{(\ell, k, r): \, \ell < k < r \\ (\ell, r) \in \mc R_{s, e}}} \frac{\left\vert \mc T_{\ell, k, r} \right\vert}{\wh{\sigma}_{r - \ell}}
\quad \text{with} \quad
\left\vert \mc T_{s_\circ, k_\circ, e_\circ} \right\vert > 
\wh{\sigma}_{e_\circ - s_\circ} \cdot D
\end{align}
for some threshold $D$ and $\wh{\sigma}^2_{r - \ell}$ denoting the proposed robust estimator of the TAVC at scale $L = r - \ell$ (when $r - \ell$ is odd, we use $\wh{\sigma}^2_{r - \ell - 1}$ instead).
As in Section~\ref{sec:mosum}, a maximum scale $M$ is set so that the CUSUM statistic over any interval of length greater than $M$ is standardised using $\wh{\sigma}_{M}$.
Following the recommendation made in \cite{fryzlewicz2014wild}, we adopt the threshold 
$D = C \sqrt{2 \log(n)}$ where $C$ is a universal constant.

If $(s_\circ, k_\circ, e_\circ)$ that fulfils~\eqref{eq:wbs:est} exists,
it signals the presence of a change point so that
the data is partitioned into $\{X_t\}_{t = s + 1}^{k_\circ}$ and $\{X_t\}_{t = k_\circ + 1}^e$, and the same step of detecting and identifying a single change point
is repeated on each partition separately.
If no such $(s_\circ, k_\circ, e_\circ)$ exists, or when the user-specified minimum segment length is reached, 
then the search for change points
is terminated on $\{X_t\}_{t = s + 1}^e$.
We provide a pseudocode of the WBS2 algorithm with the robust estimator of scale-dependent TAVC in Algorithm~\ref{alg:wbs2} of Appendix~\ref{appendix:alg}.

\subsection{Extension to local stationarity}
\label{sec:nonstat:ext}

We propose a heuristic extension of the robust estimator of scale-dependent TAVC to the setting where the second-order structure of $\{\vep_t\}_{t = 1}^n$ varies smoothly over time.
Suppose that there exists an appropriately chosen window size $W$ such that $\{\vep_t\}_{t = k - \lfloor W/2 \rfloor + 1}^{k + \lfloor W/2 \rfloor}$ may be regarded as being approximately second-order stationary over all $k$. Then, we propose to perform the robust estimation described in Section~\ref{sec:robust} in a localised fashion. 

To this end, define the time-varying TAVC at scale $L$ and time $k$ by
\begin{align}
\label{eq:tavc-tvar}
\sigma^2_L (k) = \Var\l( \frac{1}{\sqrt L} \sum_{t = k-G+1}^{k+G} \vep_t \r) .
\end{align}
%This suggests that using a simple windowed estimator for the TAVC is appropriate for such time series.
For notational convenience, we set $W = N_2 L$ for some integer $N_2$, and let $N_3 = \lfloor (W-G)/G \rfloor$ be the number of blocks for window size $W$. For $k \in \{ W/2, \ldots, n - W/2 \}$,
we estimate $\sigma^2_L(k)$ by $\wh{\sigma}^2_L(k)$,  the solution of the following $M$-estimation equation
\begin{align}\label{m-est-eq2}
h_L (u, k) = \frac{1}{N_3} \sum_{j = 1}^{N_3} \phi_v \left( \xi_{j, k - W/2} - u \right) = 0
\end{align}
with $v \asymp (G/W)^{1/2}$. 
We apply a boundary extension so that $\wh{\sigma}_L^2(1) = \ldots = \wh{\sigma}_L^2(W/2)$ and $\wh{\sigma}_{L}^2(n -W/2) = \ldots = \wh{\sigma}_L^2(n)$.
The estimator of the local TAVC at time $k$ and scale $L$ is obtained analogously as that of the global TAVC at scale $L$ described in Section~\ref{sec:robust}, except that we only use the windowed data region starting at time $k - W/2$ and ending at $k + W/2$ for the estimation of the former. 
Then, the MOSUM detector $\mc T_G(k)$ and the CUSUM statistic $\mc T_{s, k, e}$ described in Sections~\ref{sec:mosum}--\ref{sec:wbs} are standardised in a time-dependent way
using $\wh{\sigma}_{2G}(k)$ and $\wh{\sigma}^2_{(e-s)}(k)$, respectively.
In practice, we observe that taking the running median of $\{\wh{\sigma}^2_L(k + b - \lfloor G/2 \rfloor )\}_{b = 0}^{G - 1}$ as an estimator of $\sigma^2_L(k)$ further improves the performance, as it `smoothes' out the local estimators and enhances the robustness to mean changes. 

\begin{figure}[H]
\centering
\includegraphics[width = 0.8\textwidth]{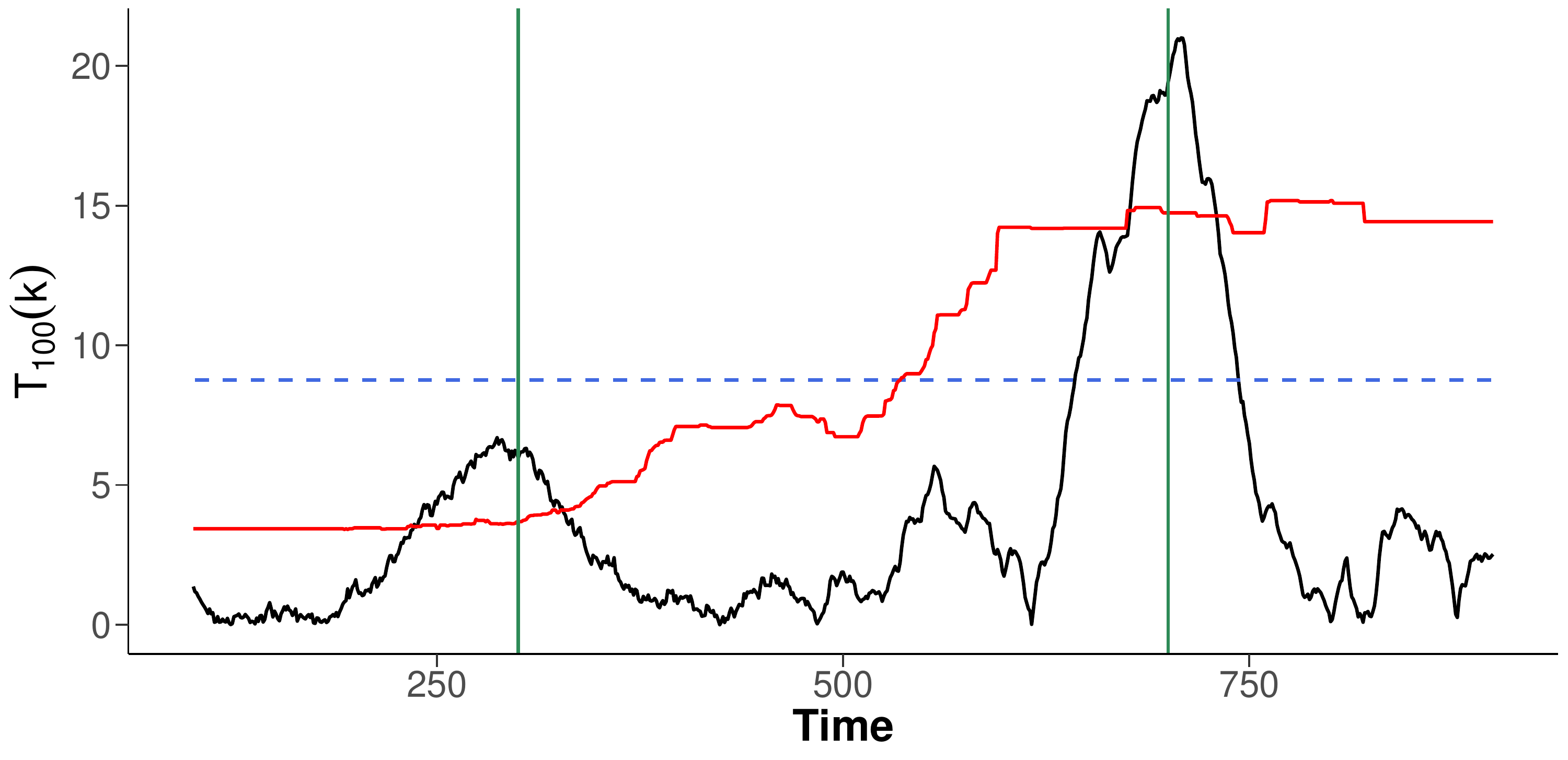}
\caption{MOSUM detector statistic (black solid line) computed from the time series generated as described in the text with two changes in the mean (denoted by vertical lines). Also shown are the thresholds computed using global (blue dashed line) and local (red solid line) estimators of the scale-dependent TAVC.}
\label{fig:toy:ex}
\end{figure}

We illustrate the benefit of adopting the time-varying adaptation of the proposed robust estimator using the following example.
Consider a time series of length $n=1000$, where the errors $\{ \varepsilon_t\}_{t \in \Z}$ follow a time-varying AR(1) model: $\varepsilon_t = a_1 (t) \varepsilon_{t-1} + W_t$, with $a_{1} (t) = 0$ for $t\leq 500$, $a_{1} (t) = 0.7$ for $t \ge 501$, and $W_t \sim_{\text{i.i.d.}} \mc N(0, 1)$. There are two changes in the mean at $\tau_1 = 300$ and $\tau_2 = 700$, with change sizes~$1$ and~$2$, respectively. 
In Figure~\ref{fig:toy:ex}, we show the MOSUM detector statistic in~\eqref{eq:mosum} calculated using bandwidth $G = 100$. 
We also plot the threshold $D_n(G, \alpha)$ at the significance level $\alpha = 0.05$, multiplied by the square root of the global TAVC estimator at scale $L = 2G$ (i.e.\ $\wh\sigma^2_{2G}$) in dashed blue line,
and that multiplied by the square root of the local estimators of the scale-$L$ TAVC (i.e.\ $\wh\sigma^2_{2G}(k)$) in solid red line. 
% Note that standardising the test statistic is equivalent to multiplying the detection threshold by $\wh{\sigma}_{2G}$. 
We see that using the global approach misses the change at time $\tau_1 = 300$ due to the global scale-$L$ TAVC estimator being too large, whilst the localised approach successfully detects both changes.

Lastly, we mention that the robust estimation of time-varying and scale-dependent TAVC is of independent interest beyond the context of change point analysis, with possible extensions including the estimation of other second-order properties. For example, the procedure can be used to obtain a robust estimator of the spectrum of a locally stationary wavelet process \citep{nason2000wavelet} while the time series undergoes shifts in the mean.

\section{Numerical results}
\label{sec:num}

%In this section, we carry out an extensive simulation study to assess the performance of the proposed method, and analyse two data examples that highlight the potential benefits of our approach.

\subsection{Practical considerations}
\label{sec:tuning}

We provide guidance on the selection of tuning parameters 
required for the proposed robust TAVC estimator and its application
with multiscale change point detection methods.
% based on extensive numerical experiments.

\vspace{-10pt}
\paragraph{Parameter $v$ in~\eqref{eq:m:est}.} 
We select $v = v_b = \bar{\xi}_b^{-1}\sqrt{G/n}$ where $\bar{\xi}_b$ is a fixed constant for a given $b \in \{0, \ldots, G - 1\}$. 
For the problem of robust mean estimation, \cite{catoni2012challenging} recommends the standard deviation in the place of $\bar{\xi}_b$
and in a similar vein, \cite{chen2021inference} propose to use a trimmed mean 
\begin{align}
\label{eq:v:one}
\bar{\xi}_{[1], b} = {\frac{1}{\lfloor 3N_1(b)/4 \rfloor - \lceil N_1(b)/4 \rceil + 1}} \sum_{j = \lceil N_1(b)/4 \rceil}^{ \lfloor 3N_1(b)/4 \rfloor} \xi_{(j), b},
\end{align}
where $\xi_{(1), b} \le \ldots \le \xi_{(N_1(b)), b}$ are the ordered $\{\xi_{j, b} \}_{j=1}^{N_1(b)}$. Another approach is to use an appropriately scaled median of $\{\xi_{j, b} \}_{j=1}^{N_1(b)}$, in a similar fashion to \cite{mcgonigle2021detecting}. In this case, the necessary scaling constant to ensure unbiasedness can be chosen by noting that $\bar{X}_{j, b}$ is asymptotically Gaussian as $L \to \infty$, and thus $\xi_{j, b}$ is asymptotically scaled $\chi^2_1$. This leads to the choice 
\begin{align}
\label{eq:v:two}
\bar{\xi}_{[2], b} = K \cdot \text{Median}\l\{ \xi_{1,b}, \ldots , \xi_{N_1(b),b} \r\},
\end{align}
where $K = 2.125$. In simulation studies, we report the results obtained with $\bar{\xi}_{[\ell], b}$, $\ell = 1, 2$, in setting the parameter $v_b$ where we observe that both choices return similarly good results.

\vspace{-10pt}
\paragraph{Parameters $b$ and $b_{\max}$ in~\eqref{eq:m:est}.}
As a further step to ensure greater robustness of the estimator, we obtain the estimator $\wh\sigma^2_{L, b}$ for a range of $b \in \{0, \ldots, b_{\max}\}$ and take their median as the final estimator $\wh\sigma^2_L$. Informally, we may get unlucky with some starting value $b$ that leads to many of the $\xi_{j, b}$  contaminated by the mean changes, and taking the median over a range of values of $b$ helps in alleviating this. % When $b_{\max}$ is fixed, this results in a consistent estimator of $\sigma^2_{L}$ as it is the median of $b_{\max}$ consistent estimators.
We take $b_{\max} = G - 1$ in practice which yields good performance. 

\vspace{-10pt}
\paragraph{Maximum time-scale $M$.} 
We recommend $M = \lfloor 2.5 \sqrt{n} \rfloor$ as the coarsest scale at which the scale-dependent TAVC is estimated. This choice is made to balance between mitigating the effect of change points, and ensuring that the TAVC at coarser scales is well-approximated by $\sigma^2_M$. 

\vspace{-10pt}
\paragraph{Tuning parameters for the multiscale MOSUM procedure.}
We follow \cite{cho2022two} and generate $\mc G$ as a sequence of Fibonacci numbers. For the simulation studies reported in Section~\ref{sec:sim} and Appendix~\ref{sec:add:sim}, we consider $\mc G = \{G_m, \, 1 \le m \le 4: \, G_1 < \ldots < G_4\}$ where $G_m = G_{m - 1} + G_{m - 2}$ for $m \ge 2$ with
$G_0 = G_1 = 20 + 10 \lfloor n/1000 \rfloor$.
 %Setting the finest bandwidth $G_1 = G_0$, we iteratively set $G_{m} = G_{m-1} + G_{m-2}$ for $m \geq 2$ until for some $H$, we have $G_{H} < G_{\max}$ and $G_{H+1} > G_{\max}$ where $G_{\max}$ is the maximum bandwidth. 
For other tuning parameters, we adopt the recommended default values of the \verb!R! package \verb!mosum! \citep{meier2021mosum}, and set $\alpha = 0.05$ and $\eta = 0.4$.

\vspace{-10pt}
\paragraph{Tuning parameters for the WBS2 algorithm.}
For the threshold, we set the constant $C = 1.3$ and draw $R = 100$ deterministic intervals at each iteration, which are suggested choices in \cite{fryzlewicz2014wild} and \cite{cho2021multiple} respectively.
As we permit the presence of serial correlations, 
it is reasonable to impose a minimum length requirement
on the intervals considered in the WBS2 algorithm.
We set this minimum length to be $2 G_1$, with $G_1$ the finest scale considered by the MOSUM procedure.

\vspace{-10pt}
\paragraph{Window size $W$ for time-varying TAVC estimation.} 
We utilise a scale-dependent window size $W_{L}$. Setting $W_{L}= N_2 L$ gives $N_3 = 2N_2-1$ data points used in the solving of the $M$-estimation equations. We advise setting $N_2 = 5$, which ensures that the influence of change points is negated and that the window size is large enough to include enough data points for reliable estimation of the TAVC.

\subsection{Simulation study}
\label{sec:sim}

In this section, we evaluate the performance of the proposed robust estimator of scale-dependent TAVC applied with the two multiscale change point detection procedures discussed in Sections~\ref{sec:mosum}--\ref{sec:wbs}. We compare with other methods that account for serial dependence under~\eqref{eq:model} and whose implementations are readily available in \verb!R!,
with a variety of scenarios for generating serially correlated $\{\vep_t\}_{t \in \Z}$.

\subsubsection{Settings}

We assess the performance of different methods both in the case of no changes ($q = 0$) and multiple changes ($q \ge 1$), under a variety of error structures. 
Unless stated otherwise, we generate 
$W_t \sim_{\text{i.i.d.}} \mc N(0, \sigma_w^2)$ with $\sigma_w = 1$.
\begin{enumerate}[label = (M\arabic*)] 
\item \label{model-a} $\varepsilon_t = W_t$.
\item \label{model-b} $\varepsilon_t = W_t$, where $W_t$ are i.i.d.\ $t_{5}$-distributed random variables.
\item \label{model-c} AR(1) model: $\varepsilon_t = a_1 \varepsilon_{t-1}+W_t$, with $a_1 = 0.9$ and $\sigma_w = \sqrt{1-a_1^2}$.
\item  \label{model-d} AR(2) model: $\varepsilon_t = a_1 \varepsilon_{t-1}+ a_2 \varepsilon_{t-2}+W_t$, with $a_1 = 0.5$ and $a_2=0.3$, with $\sigma_w = 0.6676184$.
\item \label{model-e} MA(1) model: $\varepsilon_t = W_t + b_1 W_{t-1}$, with $b_1 = -0.9$.
\item \label{model-f} ARCH(1) model: $\varepsilon_t = \sigma_t W_t$ with $\sigma_t^2 = c_0 + c_1 \varepsilon_{t-1}^2$,  where $c_0= 0.5$, $c_1 = 0.4$.
\item \label{model-g} Time-varying AR(1) model: $\varepsilon_t = a_1 (t) \varepsilon_{t-1}+W_t$, with $a_{1} (t) = 0.8-0.6t/n$.
\item  \label{model-h} Time-varying AR(1) model: $\varepsilon_t = a_1 (t) \varepsilon_{t-1}+ \sigma (t) W_t$, with $a_{1} (t) = 0.5 \cos (2\pi t/n)$ and $\sigma (t) = \sqrt{1-a_1 (t)^2}$.
\item \label{model-i} Time-varying MA(1) model: $\varepsilon_t = W_t + b_1 (t) W_{t-1}$, with $b_1 (t) = 12(t/n)^3 -18(t/n)^2+6t/n$.
\end{enumerate}

Models~\ref{model-a}--\ref{model-f} represent stationary error scenarios.
Model~\ref{model-a} is the setting commonly adopted in the literature while \ref{model-b}, taken from \cite{cho2021data}, is adopted to examine whether a method works well in the presence of non-Gaussian errors. 
Models~\ref{model-c} and~\ref{model-d} allow strong autocorrelations in $\{\vep_t\}_{t = 1}^n$. Under Model~\ref{model-e}, the LRV is close to zero, which makes its accurate estimation difficult. Model~\ref{model-f} is a non-linear process. Models~\ref{model-g}--\ref{model-i} consider time-varying dependence structure; variants of \ref{model-g} and \ref{model-h} were studied in \cite{mcgonigle2021detecting} and \cite{cho2021multiple}, respectively.
For Models~\ref{model-a}--\ref{model-f}, we use the global scale-dependent TAVC estimator described in Section~\ref{sec:robust} while for~\ref{model-g}--\ref{model-i}, we use the window-based estimator of the local scale-dependent TAVC described in Section~\ref{sec:nonstat:ext}.

We assess the performance of the methods both when $q = 0$ and $q \ge 1$.
In the latter case, the time series contains the $q$ change points at $\tau_i = \lfloor n/(q + 1) \cdot i \rfloor, \, i = 1, \ldots, q$, with the (signed) change size $\mu_i = \mu(\tau_i) \cdot (-1)^{i + 1}$.
In~\ref{model-a}--\ref{model-d} and \ref{model-f}, we set $\mu(\tau_i) = \sigma$
and in the case of~\ref{model-e}, we set $\mu = 1$.
In~\ref{model-g}--\ref{model-i}, we set $\mu(\tau_i) = \sigma(\tau_i)$
where $\sigma^2(t)$ denotes the time-varying LRV.

We implement the robust TAVC estimation within both the multiscale MOSUM and WBS2 procedures as described in Sections~\ref{sec:mosum} and~\ref{sec:wbs},
which are referred to as MOSUM.TAVC$_{[\ell]}$ and WBS2.TAVC$_{[\ell]}$, respectively.
Here, the subscript with $\ell = 1, 2$, refers to the choice of the tuning parameter $\bar{\xi}_{[\ell], b}$ involved in the parameter $v$,
see~\eqref{eq:v:one}--\eqref{eq:v:two}. 
For the choice of the tuning parameters, we refer to Section~\ref{sec:tuning}. 
For illustrative purposes, we also report the results of the `oracle' versions of the MOSUM and WBS2 procedures referred to as MOSUM.oracle and WBS2.oracle, respectively. 
These methods are implemented with the true LRV $\sigma^2$ 
($\sigma^2(t)$ in the case of Models~\ref{model-g}--\ref{model-i})
for standardising the detector statistics, 
while all other tuning parameters are kept the same.

Additionally, we consider DepSMUCE \citep{dette2020multiscale}, DeCAFS \citep{romano2021detecting} and WCM.gSa \citep{cho2021multiple} for comparison.
DepSMUCE extends SMUCE \citep{frick2014multiscale} to dependent data using a difference-type estimator of the LRV. Although not its primary objective, DeCAFS detects multiple change points in the mean assuming that the noise is a stationary AR($1$) process. The WCM.gSa method performs model selection on the sequence of models generated by the WBS2 algorithm, using an information criterion-based model selection strategy which assumes that $\{\vep_t\}_{t \in \Z}$ follows an AR model of an arbitrary order. For DepSMUCE, we consider significance levels $\alpha \in \{ 0.05, 0.2 \}$. Other tuning parameters not mentioned here are chosen as recommended by the authors. 

\subsubsection{Results}

Table~\ref{table:sim:results} summarises the results of the comparative simulation study
from $1000$ replications of time series of length $n = 1000$
generated as in~\ref{model-a}--\ref{model-i} with $q \in \{0, 4\}$.
Results for other values of $n \in \{500, 2000\}$ 
are given in Appendix~\ref{sec:add:sim},
{where we make similar observations as below.}

\begin{figure}[H]
\centering
\includegraphics[width =\textwidth]{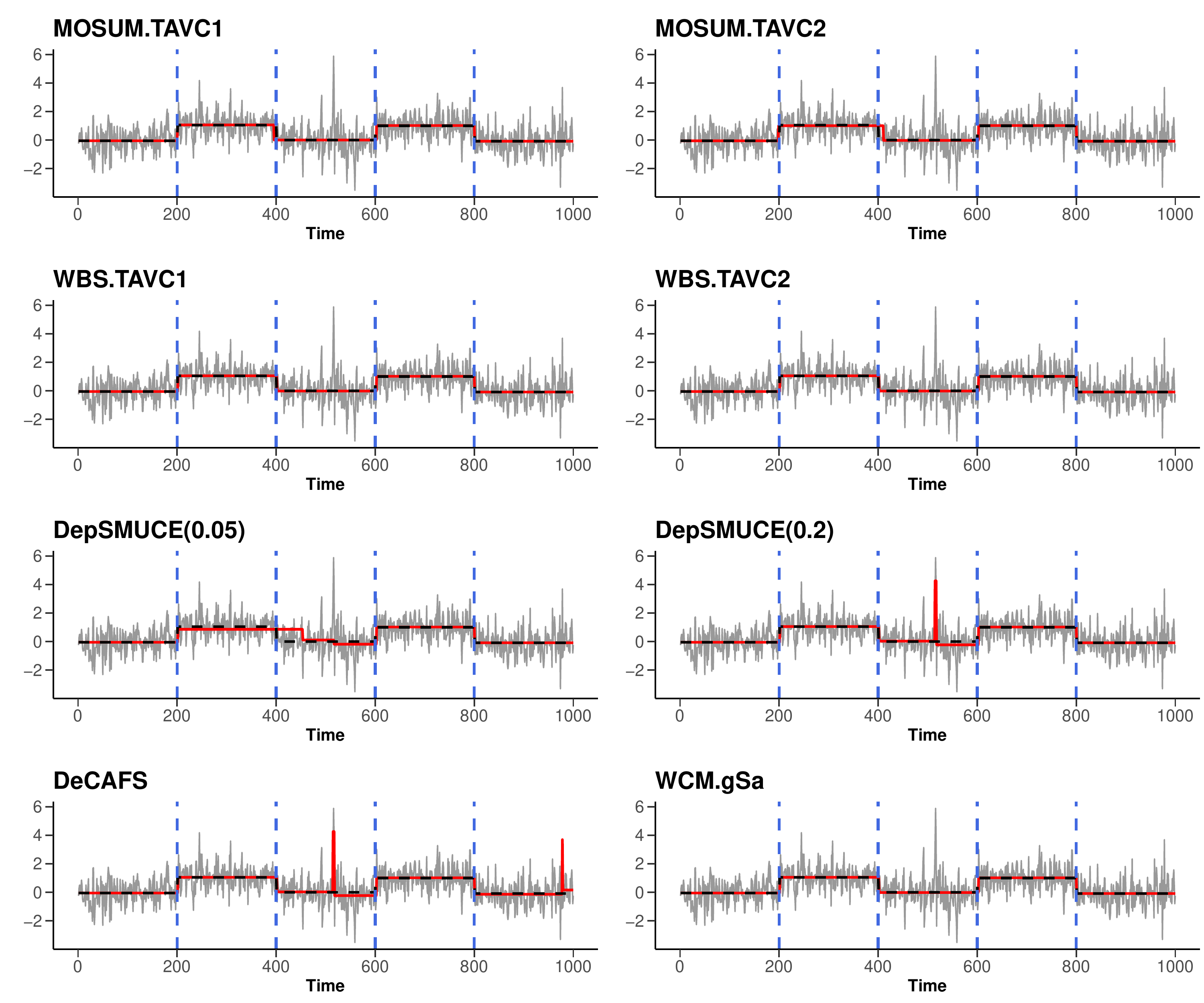}
\caption{\tcr{Comparisons of the estimator $\wh{f}_t$ (red solid line) obtained from various methods in consideration and the oracle estimator $\wh{f}^*_t$ (black dashed line) for a realisation (grey solid line) from Model~\ref{model-f}.
True change point locations are given by blue dashed vertical lines.}}
\label{fig:f-hat-plots}
\end{figure}

When $q = 0$, we report the proportion of falsely detecting any change point out of the $1000$ realisations (see the column `Size' in Table~\ref{table:sim:results}).
When $q \ge 1$, we report the relative mean squared error (RMSE)
\begin{align*}
\sum_{t=1}^{n} (\wh{f}_{t} - f_{t})^{2} / \sum_{t=1}^{n} (\wh{f}^{*}_{t} - f_{t})^{2},
\end{align*}
where $\wh{f}_{t}$ is the piecewise constant signal constructed with the estimated change points, and {$\wh{f}^*_t$} is the oracle estimator constructed with the true change points. 
\tcr{For illustration, Figure~\ref{fig:f-hat-plots} plots $\wh{f}_t$ obtained from different methods in consideration and the oracle estimator $\wh{f}^*_t$, on a realisation from Model~\ref{model-f}.} 
We also report the distribution of the estimated number of change points, as well as the covering metric (CM). The covering metric \citep{arbelaez2010contour} measures the quality of the resulting segmentation as defined by the location of the detected changes, and is recommended in \cite{van2020evaluation} as an evaluation metric for comparing change point detection algorithms. The CM take values between $0$ and $1$, with a value of $1$ corresponding to a perfect segmentation. Its explicit definition can be found in Appendix~\ref{sec:add:sim}. 
For each measure, we report its average over the $1000$ realisations.

\begingroup
{\small
\setlength{\tabcolsep}{3pt}
\setlength{\LTcapwidth}{\textwidth}
\begin{longtable}{c c c ccccc cc}
\caption{Performance comparisons for $n = 1000$. We report the size, the proportion of realisations where change points are falsely detected when $q = 0$,
and the distribution of the estimated number of change points, covering metric (CM) and relative MSE (RMSE) over 1000 realisations when $q = 4$.
The modal value of the distribution of the number of estimated change points for each method is shown in bold. The best performing method according to each metric when $q = 4$ is underlined.}
\label{table:sim:results}
\endfirsthead
\endhead
\toprule	
&&& \multicolumn{5}{c}{$\wh{q} - q$} &   &   \\ 
 Model       & Method   & Size    &  $\leq-2$     & $-1$        & $\mathbf{0}$  & $1$    & $\geq 2$    & CM & RMSE       \\ 
\cmidrule(lr){1-2} \cmidrule(lr){3-3}  \cmidrule(lr){4-8} \cmidrule(lr){9-10}
\ref{model-a}     &  MOSUM.TAVC$_{[1]}$   & 0.135  &  0.000     & 0.006   &  \bf{0.980}   & 0.014     & 0.000             & 0.967   & 6.098  \\
 &  MOSUM.TAVC$_{[2]}$  &  0.091    & 0.000    &  0.015   & \bf{0.978}  & 0.007      & 0.000        & 0.965 & 6.234 \\
 & WBS2.TAVC$_{[1]}$ & 0.049 & 0.000 & 0.004 & {\bf{0.996}} & 0.000 & 0.000 & {0.976} & {4.605}
  \\
  & WBS2.TAVC$_{[2]}$ & 0.028  & 0.001 & 0.017 & \bf{0.982} & 0.000 & 0.000 & 0.973 & 4.787
  \\ 
\cmidrule(lr){3-3}  \cmidrule(lr){4-8} \cmidrule(lr){9-10}  
       &  DepSMUCE(0.05) &  0.010   & 0.000   &  0.014    &  \bf{0.986} & 0.000      &  0.000         & 0.972  & 4.859 \\ 
     &  DepSMUCE(0.2) & 0.066    & 0.000   &  0.001    &  {\bf{0.998}} & 0.001      &  0.000         & 0.976  & \underline{4.566} \\ 
     &  DeCAFS &  0.015  & 0.000   &  0.000    &  \bf{0.970} & 0.029      &  0.001         & {0.976}  & 4.798 \\ 
     &  WCM.gSa & 0.007    & 0.000   &  0.000    &  \bf{0.978} & 0.020     &  0.002         & 0.975  & {4.733} \\
     & MOSUM.oracle & 0.040 & 0.000 & 0.000 & \bf{0.861} & 0.132 &  0.007 & 0.965 & 5.782 \\
     & WBS2.oracle & 0.004 & 0.000 & 0.000 & \underline{\bf{1.000}} & 0.000 & 0.000 & \underline{0.977} & {4.567} \\ \cmidrule(lr){1-2} \cmidrule(lr){3-3}  \cmidrule(lr){4-8} \cmidrule(lr){9-10}
     
\ref{model-b}     &  MOSUM.TAVC$_{[1]}$  &  0.149  &  0.001     & 0.004   &  \bf{0.979}   & 0.015     & 0.001             & 0.967   & 6.512  \\
 &  MOSUM.TAVC$_{[2]}$ & 0.086      & 0.003    &  0.008   & \bf{0.981}  & 0.008      & 0.000 & 0.965 & 6.595 \\
 & WBS2.TAVC$_{[1]}$ & 0.040 & 0.001  & 0.006 & {\bf{0.993}} & 0.000 & 0.000 & {0.976}
 & {4.702} \\
  & WBS2.TAVC$_{[2]}$ & 0.014 & 0.004 &  0.011 & \bf{0.985} & 0.000 & 0.000 & 0.974 & 4.899
  \\ 
\cmidrule(lr){3-3}  \cmidrule(lr){4-8} \cmidrule(lr){9-10}
       &  DepSMUCE(0.05) & 0.586     & 0.000   &  0.004    &  \bf{0.611} & 0.161      &  0.224         & 0.946  & 13.784 \\ 
    &  DepSMUCE(0.2)&  0.747    & 0.000   &  0.000    &  \bf{0.420} & 0.180      &  0.400         & 0.934  & 15.828 \\ 
    &  DeCAFS & 0.898   & 0.000   &  0.000    &  0.105 & 0.040      &  \bf{0.855}         & 0.886  & 29.258 \\ 
     &  WCM.gSa &  0.009   & 0.000   &  0.000    &  {\bf{0.978}} & 0.020      & 0.002         &{0.975}  & {4.789} \\  
     & MOSUM.oracle & 0.037 & 0.000 & 0.000 & \bf{0.843} & 0.143 & 0.014 & 0.964 & 6.300 \\
     & WBS2.oracle & 0.012 & 0.000 & 0.000  & \underline{\bf{0.996}} & 0.004 & 0.000 &  \underline{0.978} & \underline{4.650} \\ 
\cmidrule(lr){1-2} \cmidrule(lr){3-3}  \cmidrule(lr){4-8} \cmidrule(lr){9-10}
\ref{model-c}     &  MOSUM.TAVC$_{[1]}$ &  0.147   &  0.000     & 0.000   &  \bf{0.998}   & 0.002     & 0.000             & 0.995  & 1.810  \\
 &  MOSUM.TAVC$_{[2]}$ & 0.082       & 0.000    &  0.001   & \bf{0.999}  & 0.000      & 0.000        & 0.994 & 1.789 \\
 & WBS2.TAVC$_{[1]}$ & 0.062 & 0.000 & 0.000 & \underline{\bf{1.000}} & 0.000 & 0.000 & \underline{0.998} & 1.258 \\
  & WBS2.TAVC$_{[2]}$ & 0.034 & 0.000 & 0.001 & {\bf{0.999}} & 0.000 & 0.000 & \underline{0.998} & 1.264 
  \\ 
\cmidrule(lr){3-3}  \cmidrule(lr){4-8} \cmidrule(lr){9-10}
       &  DepSMUCE(0.05) & 0.968    & 0.000   &  0.000    &  {\bf{0.920}} & 0.078      &  0.002         & 0.992  & 1.494 \\ 
     &  DepSMUCE(0.2) & 0.996    & 0.000   &  0.000    &  \bf{0.739} & 0.232     &  0.029         & 0.978  & 1.888 \\ 
     &  DeCAFS &  0.597  & 0.000   &  0.000    &  \bf{0.590} & 0.342      &  0.068         & {0.982}  & \underline{1.236} \\ 
     &  WCM.gSa & 0.053    & 0.000   &  0.000    &  \bf{0.731} & 0.135      &  0.134         & 0.959  & 2.086 \\  
      & MOSUM.oracle & 0.001 & 0.000 & 0.000 & \bf{0.891} & 0.097 & 0.012 & 0.988 & 2.012 \\
     & WBS2.oracle & 0.000 & 0.000 & 0.000 & {\bf{0.979}} & 0.021 & 0.000  & 0.997 & 1.321 \\
      
     \cmidrule(lr){1-2} \cmidrule(lr){3-3}  \cmidrule(lr){4-8} \cmidrule(lr){9-10}
\ref{model-d}     &  MOSUM.TAVC$_{[1]}$ & 0.123     &  0.000     & 0.003   & {\bf{0.992}}   & 0.005     & 0.000              & {0.987}   & 3.133  \\
 &  MOSUM.TAVC$_{[2]}$ & 0.073       & 0.001    &  0.004   & \bf{0.992}  & 0.003      & 0.000        & 0.986 & 3.232 \\
 & WBS2.TAVC$_{[1]}$ & 0.053 & 0.001 & 0.000 & {\bf{0.999}} & 0.000 & 0.000 & {0.994}
 & {1.715} \\
  & WBS2.TAVC$_{[2]}$ & 0.035 &0.003 & 0.002& \bf{0.995} & 0.000 & 0.000 & {0.993} & 1.755
  \\ 
\cmidrule(lr){3-3}  \cmidrule(lr){4-8} \cmidrule(lr){9-10}  
       &  DepSMUCE(0.05) & 0.678    & 0.000   &  0.000    &  \bf{0.979} & 0.021      &  0.000  & {0.993}  & 1.827 \\ 
     &  DepSMUCE(0.2) & 0.907     & 0.000   &  0.000    &  \bf{0.891} & 0.105      &  0.004         & 0.986  & 2.089 \\ 
    &  DeCAFS &  0.734  & 0.000   &  0.000    &  0.241 & 0.224      &  \bf{0.535}         & 0.935  & {2.258} \\ 
     &  WCM.gSa &  0.022   & 0.000   &  0.000    &  \bf{0.778} & 0.111      &  0.111         & 0.966  & 2.684 \\  
      & MOSUM.oracle & 0.008 & 0.000 & 0.000 & \bf{0.972} & 0.028 & 0.000 & 0.986 &  3.087 \\
     & WBS2.oracle & 0.001 & 0.000 & 0.000 & \underline{\bf{1.000}} & 0.000 & 0.000 & \underline{0.995} & \underline{1.705} \\ 
\cmidrule(lr){1-2} \cmidrule(lr){3-3}  \cmidrule(lr){4-8} \cmidrule(lr){9-10}     
\ref{model-e}     &  MOSUM.TAVC$_{[1]}$ & 0.120    &  0.000     & 0.000   &  \underline{\bf{1.000}}   & 0.000     & 0.000             & 0.990   & 89.580  \\
 &  MOSUM.TAVC$_{[2]}$ &  0.069      & 0.000    &  0.000   & \underline{\bf{1.000}}  & 0.000      & 0.000        & 0.990 & 89.580 \\
 & WBS2.TAVC$_{[1]}$ & 0.103 & 0.000 & 0.000 & \underline{\bf{1.000}} & 0.000 & 0.000 & \underline{0.992} & 76.922 \\
  & WBS2.TAVC$_{[2]}$ & 0.052 & 0.000 & 0.000 & \underline{\bf{1.000}} & 0.000 & 0.000 & \underline{0.992} & 76.922 
  \\ 
\cmidrule(lr){3-3}  \cmidrule(lr){4-8} \cmidrule(lr){9-10}
       &  DepSMUCE(0.05)&  0.998    & 0.000   &  0.000    &  0.036 & 0.048      &  \bf{0.916}         & 0.773  & 2535.266 \\ 
     &  DepSMUCE(0.2) &1.000     & 0.000   &  0.000    &  0.003 & 0.004      &  \bf{0.993}         & 0.634  &1038.395 \\ 
     &  DeCAFS  & 0.001  & 0.000   &  0.000    &  \bf{0.997} & 0.003      &  0.000         & \underline{0.992}  & 77.886 \\ 
     &  WCM.gSa & 0.000    & 0.000   &  0.000    &  \underline{\bf{1.000}} & 0.000      &  0.000         & \underline{0.992}  & \underline{76.814}  \\ 
 & MOSUM.oracle & 1.000 & 0.000 & 0.000 & 0.000 & 0.000 & \bf{1.000} & 0.276 & 247.634 \\
     & WBS2.oracle & 1.000 & 0.000 & 0.000 & 0.000 & 0.000 & \bf{1.000} & 0.278 & 207.555 \\ 
\cmidrule(lr){1-2} \cmidrule(lr){3-3}  \cmidrule(lr){4-8} \cmidrule(lr){9-10}
\ref{model-f}     &  MOSUM.TAVC$_{[1]}$ & 0.168    &  0.000     & 0.000   &  \bf{0.978}   & 0.021     & 0.001             & 0.973   & 6.246  \\
 &  MOSUM.TAVC$_{[2]}$ &  0.112     & 0.000    &  0.001   & {\bf{0.993}}  & 0.006      & 0.000        & 0.973 & 6.246 \\
 & WBS2.TAVC$_{[1]}$ & 0.064 & 0.000 & 0.000 & \underline{\bf{1.000}} & 0.000 & 0.000 & \underline{0.981} & \underline{4.833} \\
  & WBS2.TAVC$_{[2]}$ & 0.030 & 0.000 & 0.001  & {\bf{0.999}} & 0.000 & 0.000 & \underline{0.981} & {4.844}
  \\ 
\cmidrule(lr){3-3}  \cmidrule(lr){4-8} \cmidrule(lr){9-10}
       &  DepSMUCE(0.05) & 0.507     & 0.000   &  0.000    &  \bf{0.740} & 0.176      &  0.084         & 0.963  & 8.026 \\ 
     &  DepSMUCE(0.2) &  0.716    & 0.000   &  0.000    &  \bf{0.564} & 0.252      &  0.184         & 0.949  & 9.934 \\ 
     &  DeCAFS & 0.755   & 0.000   &  0.000    &  0.191 & 0.066      &  \bf{0.743}         & 0.911  & 25.250 \\ 
     &  WCM.gSa & 0.021    & 0.000   &  0.000    &  \bf{0.971} & 0.018      &  0.011         & {0.978}  & {5.406} \\ 
      & MOSUM.oracle & 0.021 & 0.000 & 0.000 & \bf{0.881} & 0.109 & 0.010 & 0.969 & 6.300 \\
     & WBS2.oracle & 0.005 & 0.000 & 0.000 & \bf{0.995} & 0.005  & 0.000 & \underline{0.981} & {4.863} \\ 
\cmidrule(lr){1-2} \cmidrule(lr){3-3}  \cmidrule(lr){4-8} \cmidrule(lr){9-10}
\ref{model-g}  &  MOSUM.TAVC$_{[1]}$ & 0.247     & 0.000    &  0.002   & \bf{0.970}  & 0.025      & 0.003        & 0.973 & 5.232 \\    &  MOSUM.TAVC$_{[2]}$ & 0.171   &  0.000     & 0.004   &  \bf{0.972}   & 0.023     & 0.001  & 0.972   & 5.310 \\
 & WBS2.TAVC$_{[1]}$ & 0.184 & 0.000 & 0.004 & \bf{0.988} & 0.008 & 0.000 & {0.971} & {5.004} \\
  & WBS2.TAVC$_{[2]}$ & 0.125 &0.000 & 0.010 & {\bf{0.987}} & 0.003 & 0.000 & 0.970 & 5.114
  \\ 
\cmidrule(lr){3-3}  \cmidrule(lr){4-8} \cmidrule(lr){9-10}
       &  DepSMUCE(0.05) & 0.747    & 0.000   &  0.175    &  \bf{0.718} & 0.107 &  0.000  & 0.928  & 7.807 \\ 
     &  DepSMUCE(0.2) & 0.921    & 0.000   &  0.019    &  \bf{0.716} & 0.258      &  0.007  & 0.960  & 5.416 \\ 
     &  DeCAFS & 0.830   & 0.000   &  0.000 &  \bf{0.652} & 0.171      &  0.177         & 0.958  & 6.217 \\ 
     &  WCM.gSa & 0.471    & 0.059   &  0.041    &  \bf{0.830} & 0.044      &  0.026         & 0.941  & 6.975 \\ 
      & MOSUM.oracle & 0.015 & 0.000 & 0.000 & \bf{0.903} & 0.094 & 0.003 & 0.969 & 5.312 \\
     & WBS2.oracle & 0.005 & 0.000 & 0.000 & \underline{\bf{1.000}} & 0.000 & 0.000 & \underline{0.975} & \underline{4.884} \\ 
\cmidrule(lr){1-2} \cmidrule(lr){3-3}  \cmidrule(lr){4-8} \cmidrule(lr){9-10}
\ref{model-h}     &  MOSUM.TAVC$_{[1]}$  & 0.244  &  0.000     & 0.001   &  \bf{0.947}   & 0.052     & 0.000  & 0.969   & 6.884  \\
 &  MOSUM.TAVC$_{[2]}$ & 0.154     & 0.000    &  0.001   & \bf{0.961}  & 0.038 & 0.000        & 0.969 & 6.918 \\
  & WBS2.TAVC$_{[1]}$ & 0.160 & 0.000 & 0.002 &  \bf{0.995} & 0.003 & 0.000 & {0.967} & {6.565}  \\
  & WBS2.TAVC$_{[2]}$ & 0.107 & 0.000 & 0.003 & {\bf{0.994}} & 0.003 & 0.000 & 0.967 & 6.658
  \\ 
\cmidrule(lr){3-3}  \cmidrule(lr){4-8} \cmidrule(lr){9-10}
       &  DepSMUCE(0.05) & 0.219     & 0.032   &  \bf{0.641}    &  0.306 & 0.019 &  0.002  & 0.817  & 17.602 \\ 
     &  DepSMUCE(0.2) & 0.483    & 0.000   &  0.252    &  \bf{0.633} & 0.105  &  0.010 & 0.904  & 10.411 \\ 
     &  DeCAFS & 0.341     & 0.001   &  0.002    &  \bf{0.652} & 0.153      &  0.192   & 0.953  & 9.188 \\ 
     &  WCM.gSa & 0.173    & 0.076   &  0.159    &  \bf{0.748} & 0.011      &  0.006         & 0.913  & 8.976 \\
    & MOSUM.oracle & 0.045 & 0.000 & 0.000 & \bf{0.841} & 0.142 & 0.017 & 0.962 & 7.206 \\
     & WBS2.oracle & 0.013 & 0.000 & 0.000 & \underline{\bf{0.996}} & 0.004 & 0.000 & \underline{0.971} & \underline{5.960} \\ \cmidrule(lr){1-2} \cmidrule(lr){3-3}  \cmidrule(lr){4-8} \cmidrule(lr){9-10}
\ref{model-i}    &  MOSUM.TAVC$_{[1]}$  & 0.311   &  0.000     & 0.008   &  \bf{0.915}   & 0.076     & 0.001  & \underline{0.963}   & 7.027  \\
 &  MOSUM.TAVC$_{[2]}$ & 0.204      & 0.000    &  0.020   & \bf{0.931}  & 0.049 & 0.000  & 0.962 &  \underline{7.026} \\
  & WBS2.TAVC$_{[1]}$ & 0.234 & 0.000 & 0.020 & \underline{\bf{0.972}} & 0.008 & 0.000 & {0.958} & 8.451 \\
  & WBS2.TAVC$_{[2]}$ & 0.167 & 0.000 & 0.029 & \bf{0.968} & 0.003 & 0.000 & 0.958 & 8.143
  \\ 
\cmidrule(lr){3-3}  \cmidrule(lr){4-8} \cmidrule(lr){9-10}
       &  DepSMUCE(0.05) & 0.130     & 0.181   &  \bf{0.806}    &  0.013 & 0.000 &  0.000  & 0.743  & 13.448 \\ 
     &  DepSMUCE(0.2) & 0.380   & 0.022   &  \bf{0.874}    &  {0.094} & 0.010  &  0.000 & 0.774  & 11.770 \\ 
     &  DeCAFS & 0.075    & 0.034   &  0.414    &  \bf{0.431} & 0.078      &  0.043   & 0.858  & 10.028 \\ 
     &  WCM.gSa & 0.054    & 0.057   &  \bf{0.695}    &  {0.236} & 0.009     &  0.003        & 0.814  & 10.822 \\
      & MOSUM.oracle & 0.101 & 0.000 & 0.000 & \bf{0.760} & 0.211 & 0.029 & 0.957 & 7.436 \\
     & WBS2.oracle & 0.080 & 0.000 & 0.000 & \bf{0.827} & 0.171 & 0.002 & 0.950 & 7.927
     \\
\bottomrule
\end{longtable}}
\endgroup

Overall, WBS2.TAVC displays better size control than MOSUM.TAVC. 
The multiscale MOSUM procedure with bottom-up merging has been noted to return false positives as it accepts all estimators from the finest bandwidth; see the simulation results reported in \cite{cho2022two}.
Despite this known issue, MOSUM.TAVC shows better size control than some of the competitors such as DepSMUCE and DeCAFS. Between the two choices of the parameter $v$ used in~\eqref{eq:m:est},
the one involving~\eqref{eq:v:two} (corresponding to the subscript $2$) yields the estimator of TAVC 
that returns better size control, e.g.\ closer to the nominal level $\alpha = 0.05$
for the multiscale MOSUM procedure.
On the other hand, using the trimmed mean (corresponding to the subscript $1$) as in~\eqref{eq:v:one} sees improved power at the cost of larger size. This suggests that an approach combining the two choices of $v$ may yield a more balanced performance.

WBS2.TAVC performs the best across all metrics and scenarios among non-oracle methods when $q \ge 1$. 
Also, we observe that using the proposed robust estimator of scale-dependent TAVC,
compares favourably to the multiscale methods applied with the true LRV
(i.e.\ MOSUM.oracle and WBS2.oracle) and
in some scenarios, the former outperforms the respective oracle counterpart.  
In particular, in Scenario~\ref{model-e} where the LRV is close to $0$, plugging in its true value leads to detecting many false positives. 
This shows that adopting the scale-dependent TAVC in place of the LRV for test statistic standardisation improves the finite sample performance when the change point detection procedure involves localised testing, as is the case for both the MOSUM and the WBS2 procedures.
We make a similar observation about the performance of DepSMUCE which also sets out to estimate the LRV.
DeCAFS exhibits good detection power but tends to over-estimate the number of change points as well as failing to control the size adequately
even when it is applied to the correctly specified scenario (Model~\ref{model-c}).
WCM.gSa performs well in correctly estimating the number of change points
regardless of whether $q  = 0$ or $q \ge 1$. However, its performance  deteriorates in the presence of nonstationarities,
see \ref{model-g}--\ref{model-i}. 

\begin{figure}[H]
\centering
\includegraphics[width =\textwidth]{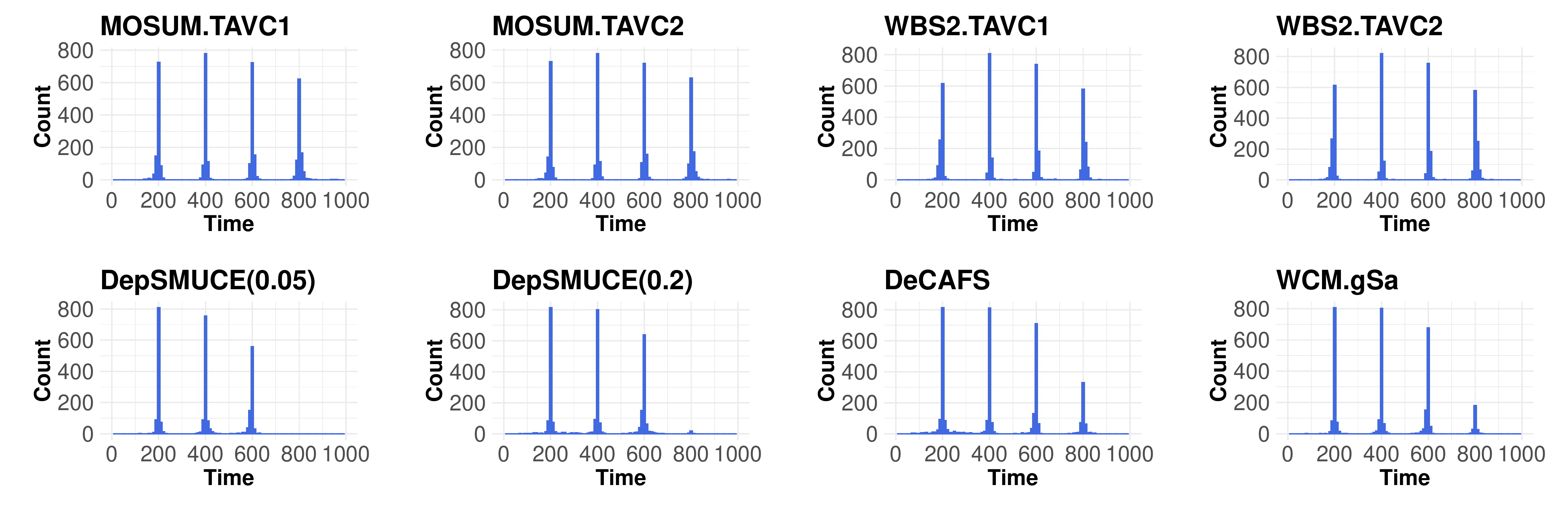}
\caption{Histograms plotting the change point estimators returned by different methods under Model~\ref{model-i}.}
\label{fig:cpts-hist}
\end{figure}

Further inspection of the results under~\ref{model-i} demonstrates one advantage of the time-varying approach. Figure~\ref{fig:cpts-hist} plots the histogram of the estimated change point locations across the $1000$ replications for each of the methods. We see that the competing approaches struggle to detect the final change point due to the decreased variability towards the end of the data sequence,
whereas the proposed estimator of time-varying scale-dependent TAVC successfully extends to the locally stationary scenarios.

\subsection{Data applications}

We apply MOSUM.TAVC and WBS2.TAVC, the multiscale procedures combined with the robust estimator of scale-dependent TAVC, to two data examples. 
We select the parameter $v$ in~\eqref{eq:m:est} using~\eqref{eq:v:one} and select other tuning parameters as described in Section~\ref{sec:tuning} unless specified otherwise.

\subsubsection{House price index data}

We analyse the monthly percentage changes in UK house price index (HPI), which provides insight into the estimated overall changes in house prices across the UK. The data are available from \url{https://www.gov.uk/government/statistical-data-sets/},
%uk-house-price-index-data-downloads-december-2021}, 
and a detailed description of the calculation of the HPI can be found from \cite{ukland}. The HPI series for various regions of the UK have previously been analysed in \cite{baranowski2019narrowest} and \cite{mcgonigle2021detecting}. We analyse the HPI for detached properties in the area of Somerset West and Taunton between April 1995 and February 2022 ($n = 323$).

\begin{figure}[H]
\centering
\includegraphics[width =0.8\textwidth]{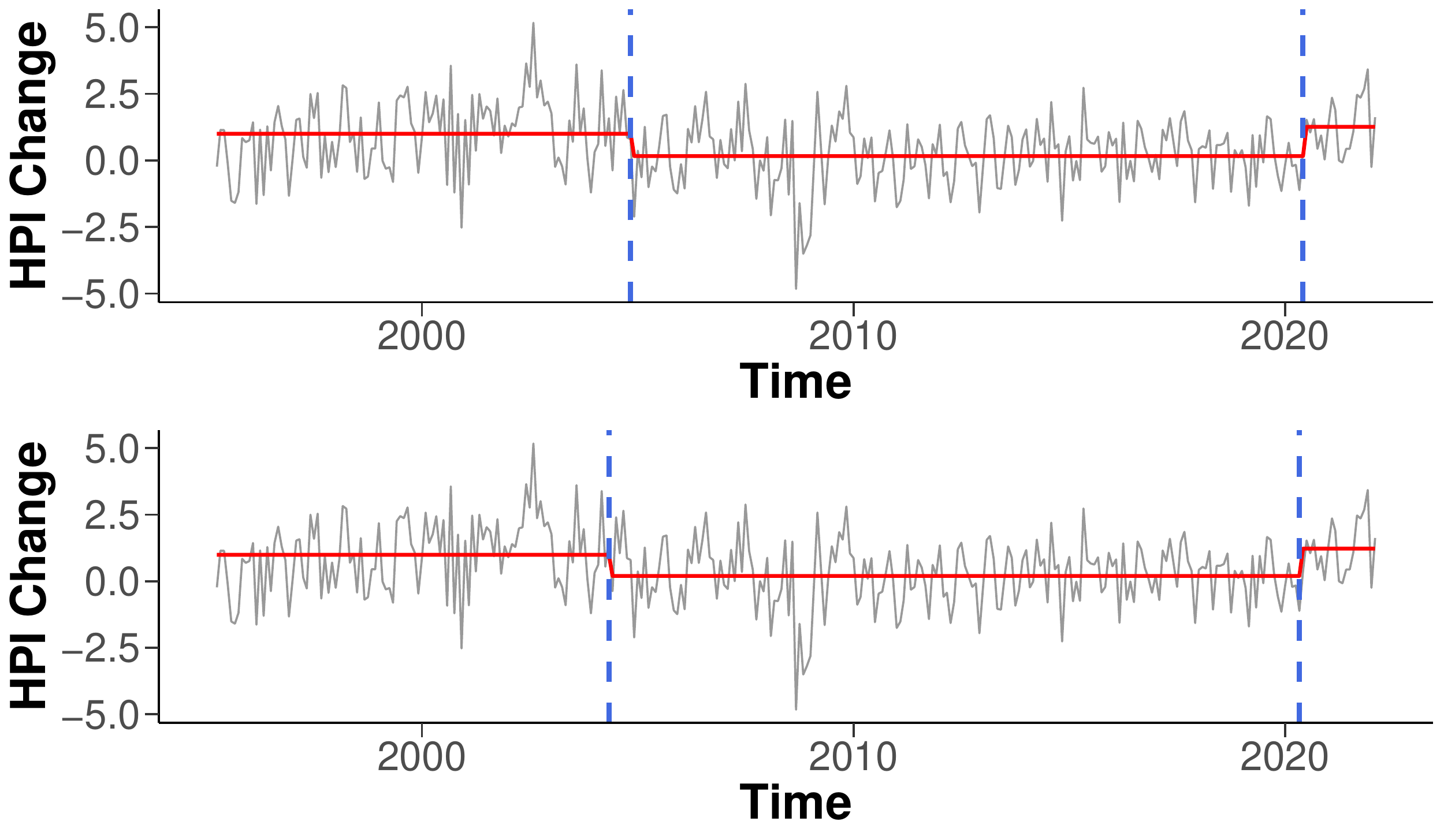}
\caption{Monthly percentage change in HPI for detached properties in Somerset West and Taunton. Change point estimators returned by WBS2.TAVC (top) and MOSUM.TAVC (bottom) are denoted by dashed vertical lines, and the estimated means are given by solid line.}
\label{fig:taunton}
\end{figure}

We set the tuning parameters as described in Section~\ref{sec:tuning}
except for the window size $W = 4L$, $C = 1.15$ (for WBS2.TAVC) and $\alpha = 0.1$ (for MOSUM.TAVC) due to the short length of the time series. We combine the multiscale change point detection procedures with the robust estimator of the time-varying, scale-dependent TAVC described in Section~\ref{sec:nonstat:ext}, with the bandwidths $\mc G = \{ 20,40,60 \}$ for the MOSUM procedure and the minimum interval length set at $40$ for WBS2. The data are shown in Figure~\ref{fig:taunton}, with the change points detected by WBS2.TAVC and MOSUM.TAVC as well as the resulting estimated mean signal given in the top and bottom panels, respectively. 
For comparison, we also apply DepSMUCE, DeCAFS and WCM.gSa to the data, see Table~\ref{table:hpi}.

Both WBS2.TAVC and MOSUM.TAVC detect two changes. The first change, detected in May and November 2004 for the two methods, corresponds to a decrease in the mean of the HPI series. The second change, detected in May/June 2020, may be associated with the changing consumer demand for housing in the wake of the COVID-19 pandemic. The Taunton area saw the biggest increase in overall house prices in 2021, as the ``race for space" saw buyers opt for ``more space to work from home as well as more outdoor space" \citep{guardian}. 

\begin{figure}[H]
\centering
\includegraphics[width =0.75\textwidth]{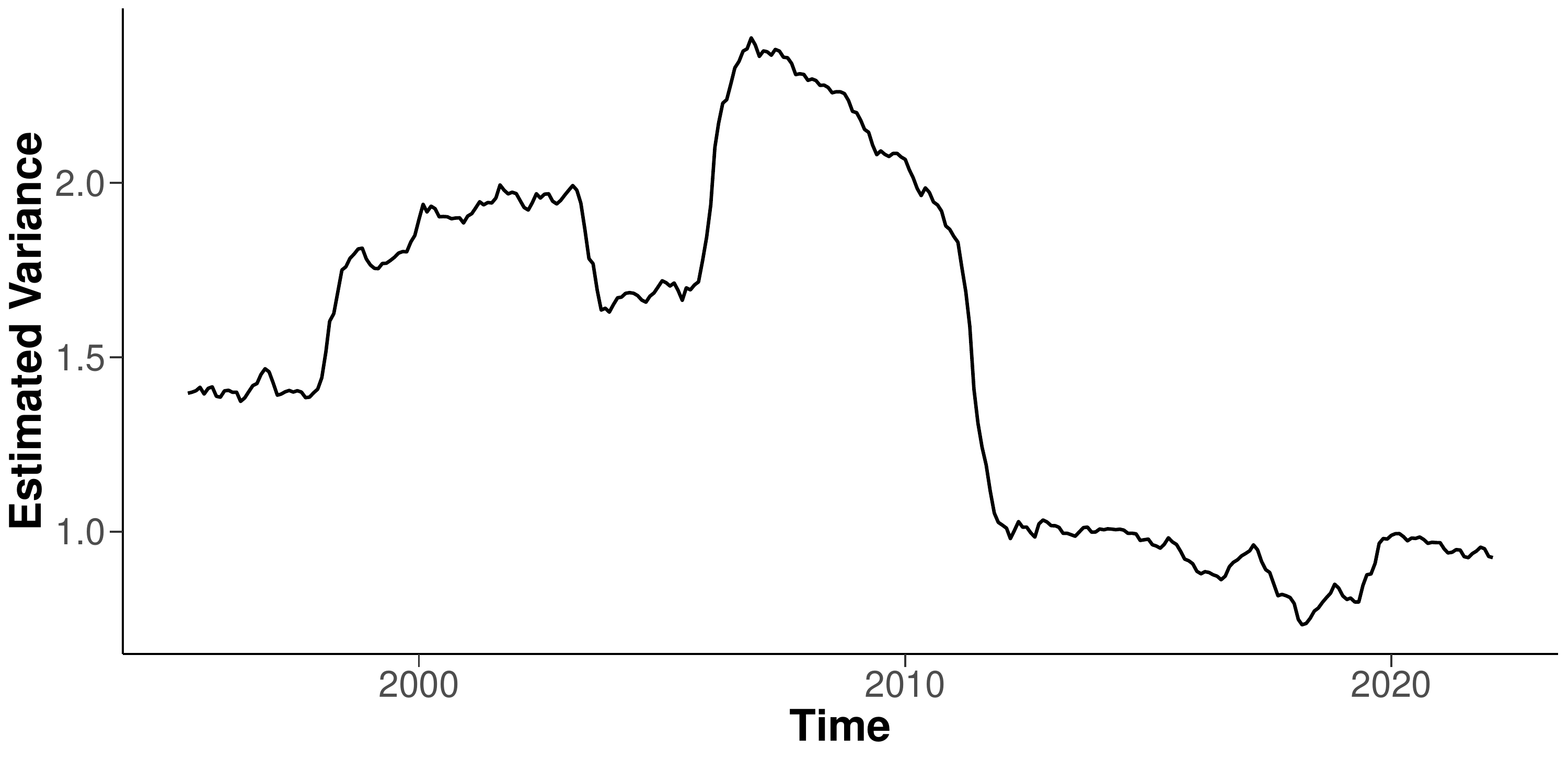}
\caption{Estimated variance of the mean change-adjusted Somerset West and Taunton HPI series.}
\label{fig:taunton:var}
\end{figure}

We observe that no changes are detected by either WBS2.TAVC or MOSUM.TAVC during the 2008--2009 period associated with the financial crisis. In contrast, DepSMUCE (with $\alpha = 0.2$), DeCAFS and WCM.gSa detect changes during this time period, possibly influenced by the increased variability during the financial crisis.
Changes detected in the crisis period can be attributed to changes in variance (and autocorrelation), rather than those in mean, as noted in \cite{mcgonigle2021detecting}. We further support this interpretation by estimating the time-varying variance after adjusting for the shifts in mean using the change point estimators returned by WBS2.TAVC, 
using the wavelet-based framework of \cite{nason2000wavelet} implemented for non-dyadic data as described in \cite{mcgonigle2022trend}, see Figure~\ref{fig:taunton:var}. There is a clear period of increased variance between 2007--2010, likely due to the financial crisis. 
By utilising a time-varying TAVC estimator, our proposed methodology is able to capture the increased variability during this period, which ensures that potential false positives are not detected. 
Furthermore, by accounting for the decrease in variability towards the end of the series, our time-varying estimator of the scale-dependent TAVC allows for the detection of a change in 2020 that is missed by other methods.

\begin{table}[htb]
\centering
\caption{Change points detected from the monthly HPI series for Somerset West and Taunton from April 1995 to February 2022.
}
\label{table:hpi}
{
\setlength{\tabcolsep}{2pt}
\begin{tabular}{ll}
\toprule
Method & Detected change points \\
\cmidrule(lr){1-1} \cmidrule(lr){2-2}
WBS2.TAVC & 2004-11, 2020-06 \\
MOSUM.TAVC & 2004-05, 2020-05 \\
\cmidrule(lr){1-1} \cmidrule(lr){2-2}
DepSMUCE(0.05) & 2004-11 \\
DepSMUCE(0.2) & 2008-09 \\
DeCAFS & 1999-05, 2003-01, 2008-08, 2009-01 \\
WCM.gSa & 1999-05, 2003-01, 2007-09, 2009-01, 2021-08 \\
\bottomrule
\end{tabular}}
\end{table}

\subsubsection{Nitrogen dioxide concentration in London}

We analyse the daily average concentrations of nitrogen dioxide (NO$_2$), measured in $\mu g/m^3$, recorded at Marylebone Road in London, UK. The measurements were taken from January 1st, 2000 until October 31st, 2021 ($n = 7734$). 
The data set is available from \url{https://www.londonair.org.uk} and a similar dataset was analysed for shifts in the mean in \cite{cho2021multiple} using the WCM.gSa method. 
The data take positive values and display both seasonality and effects due to bank holidays, as the main source of NO$_2$ emissions at the site is likely to be road traffic.  
To mitigate these effects, we take the square root transform of the data and remove seasonality as described in \cite{cho2021multiple}.

\begin{figure}[H]
\centering
\includegraphics[width =0.95\textwidth]{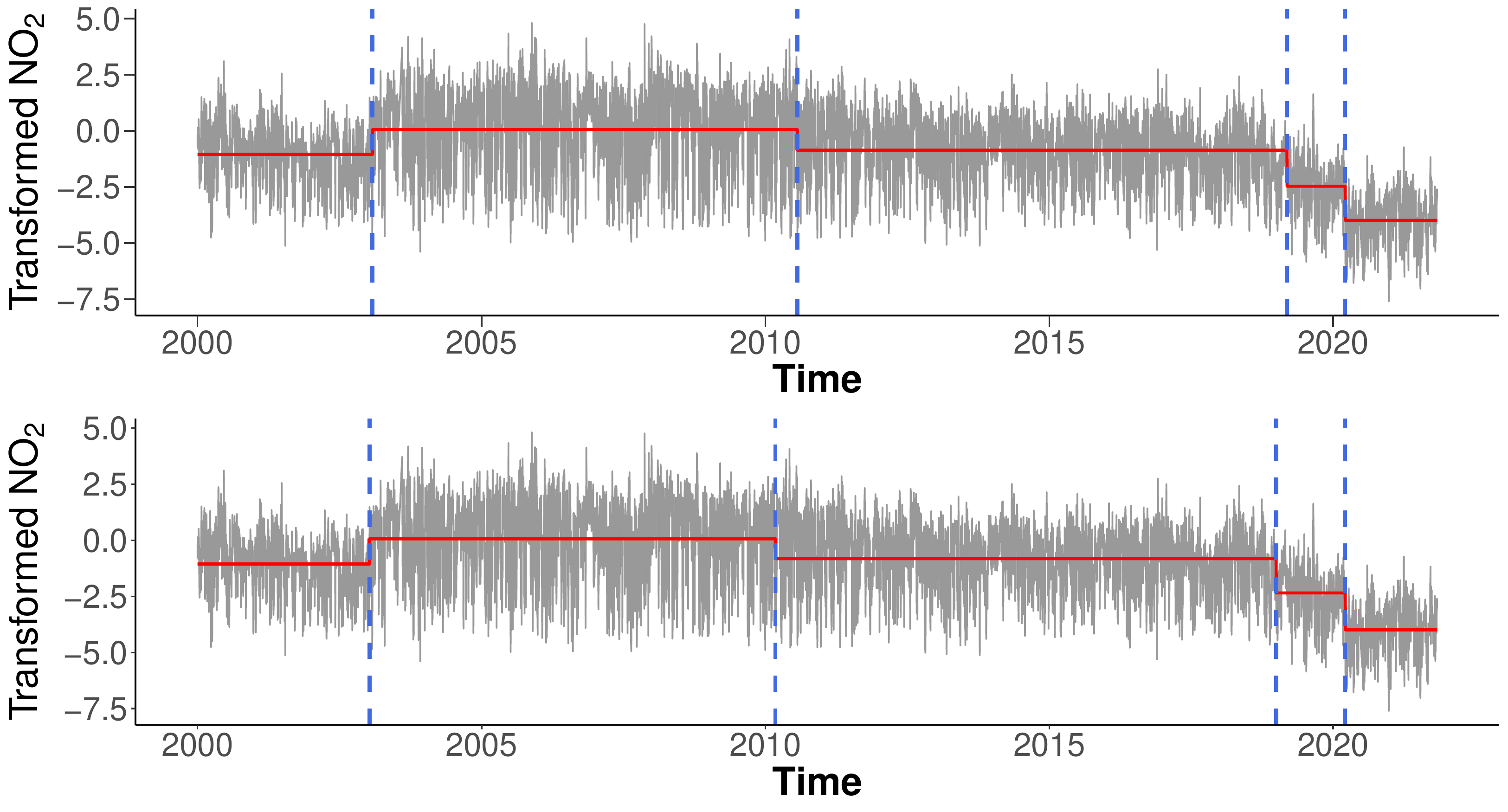}
\caption{Transformed daily NO$_2$ measurements taken at Marylebone Road, London, UK. Change point estimators returned by WBS2.TAVC (top) and MOSUM.TAVC (bottom) are denoted by dashed vertical lines, and the estimated meanx are given by solid lines.}
\label{fig:no2}
\end{figure}

We apply WBS2.TAVC and MOSUM.TAVC using the global TAVC estimator in~\eqref{eq:m:est}, with the minimum interval length set at $80$ for WBS2.TAVC and the bandwidths set as $\mc G = \{ 40,80,120,200,320,520,840 \}$ for the MOSUM.TAVC.
All other tuning parameters are selected as in Section~\ref{sec:tuning}. 
The transformed data are shown in Figure~\ref{fig:no2}, with change points detected by WBS2.TAVC and MOSUM.TAVC and the resulting estimated mean signals given in the top and bottom panels, respectively. 
For comparison, we also apply DepSMUCE, DeCAFS and WCM.gSa to the data.
Except for DeCAFS, which detects 17 change points, all methods return similar estimators. For brevity, the DeCAFS method is omitted from the results reported in Table~\ref{table:no2}.

Both WBS2.TAVC and MOSUM.TAVC detect four change points, some of which can be linked to policy changes likely affecting the levels of air pollutants. 
In February 2003, traffic management measures were introduced in central London which included modification of the pollutant filters of London buses and other heavy duty diesel vehicles, leading to an increase in their NO$_2$ emissions \citep{airquality}. 
This corresponds to the change on January 31st, 2003 detected by all methods. 
Also, Marylebone Road lies within the ultra low emission zone (ULEZ) that was introduced in April 2019. The ULEZ places restrictions on the levels of pollutants of vehicles travelling in the zone, and can be linked to the change on March 10th, 2019 detected by WBS2.TAVC or earlier in 2018 by other methods considering the bias in the change point estimators. 
This corresponds to a decrease in the concentration of NO$_2$. 
The final change point, detected by all methods on March 18th, 2020, aligns with the nationwide lockdown due to the COVID-19 pandemic on March 23rd, 2020, which resulted in drastically reduced levels of NO$_2$ throughout the UK \citep{higham2021uk}. 

\begin{table}[htb]
\centering
\caption{Change points detected from the daily average concentrations of NO$_2$ at Marylebone Road in London from January 1st, 2000 to October 31st, 2021.
}
\label{table:no2}
{
\setlength{\tabcolsep}{2pt}
\begin{tabular}{ll}
\toprule
Method & Detected change points \\
\cmidrule(lr){1-1} \cmidrule(lr){2-2}
WBS2.TAVC & 2003-01-31, 2010-07-25, 2019-03-10, 2020-03-18 \\
MOSUM.TAVC & 2003-01-11, 2010-03-06, 2018-12-30, 2020-03-18 \\
\cmidrule(lr){1-1} \cmidrule(lr){2-2}
DepSMUCE(0.05) & 2003-01-31, 2010-07-25, 2018-10-14, 2020-03-18 \\
DepSMUCE(0.2) & 2003-01-31, 2008-08-31, 2012-10-04, 2018-10-14, 2020-03-18 \\
WCM.gSa & 2003-01-31, 2009-12-09, 2018-10-14, 2020-03-18 \\
\bottomrule
\end{tabular}}
\end{table}

\section{Conclusions}
\label{sec:conc}

We propose an estimator of scale-dependent TAVC that is robust to the presence of (possibly) multiple mean shifts. It is readily combined with multiscale change point detection methodologies which, by scanning for change points over data sections of varying lengths, provide good adaptivity to the problem of multiple change point detection. We show the consistency of the proposed estimator under general assumptions permitting heavy tails and serial dependence decaying at a polynomial rate, and investigate its use with the multiscale MOSUM procedure and the WBS2 algorithm. 
Through extensive numerical studies, we demonstrate the benefit of adopting the proposed estimator of scale-dependent TAVC for improved finite sample performance, as it better reflects the level of variability in the local data sections involved in the multiscale methods.
In particular, the heuristic extension to local stationarity shows promising performance which provides a natural avenue for future research. \tcr{An implementation of the methodology in the} \verb!R! \tcr{programming language  can be found at \url{https://github.com/EuanMcGonigle/TAVC.seg}.}

\bibliographystyle{apalike}
\bibliography{research-new}

\clearpage
\appendix

\appendixpage
\numberwithin{equation}{section}
\numberwithin{figure}{section}
\numberwithin{table}{section}

\section{Algorithms and further description}
\label{appendix:alg}

\subsection{Multiscale MOSUM procedure with bottom-up merging}

Algorithm~\ref{alg:mosum} provides a pseudocode for the multiscale MOSUM procedure with bottom-up merging combined with the robust estimation of TAVC.

\begin{algorithm}[htp]
\caption{Multiscale MOSUM procedure with bottom-up merging}
\label{alg:mosum}
\DontPrintSemicolon
\SetAlgoLined
\SetKwData{return}{return}

\SetKwProg{Fn}{Function}{:}{}

\KwIn{Data $\{X_t\}_{t = 1}^n$, set $\mathcal{G}$ of bandwidths, $\alpha, \eta \in (0,1)$, maximum scale $M$ for TAVC estimation}

Initialise $\mc P \leftarrow \mc C \leftarrow \emptyset$ 
\BlankLine
\For{$G \in \mc G$}{
    \uIf{\textup{$2G \le M$}}{ %\vspace{3pt}
	Set $\wh{\sigma}^2_{2G}$ as the solution to~\eqref{eq:m:est} with $L = 2G$
	} \Else{
	    Set $\wh{\sigma}^2_{2G} \leftarrow \wh{\sigma}^2_M$
	    with $\wh{\sigma}^2_M$ solving~\eqref{eq:m:est} with $L = M$
	}
	\BlankLine
	
    $\mc C(G) \leftarrow$ Set of change point estimators obtained with bandwidth $G$ and
critical value $D_n(G, \alpha)$ according to~\eqref{eq:mosum:est} 
\BlankLine

\lFor{$\wh{k} \in \mc C(G)$}{ Add $(\wh{k},G)$ to $\mc P$}
}
\BlankLine

\For{$\wh{k}_\circ \in \mc P$ in increasing order with respect to $G$}{
    \lIf{\textup{$\min_{\wh{k} \in \mc C}| \wh{k}_\circ - \wh{k}  | \geq \eta G$}}{Add $\wh{k}_\circ$ to $\mc C$}
}
\BlankLine

\KwOut{$\mc C$}
\end{algorithm}

\subsection{Wild binary segmentation 2 algorithm}

Algorithm~\ref{alg:wbs2} provides a pseudocode for the WBS2 algorithm combined with the robust TAVC estimation. 

\begin{algorithm}[t]
\caption{Wild binary segmentation~2}
\label{alg:wbs2}
\DontPrintSemicolon
\SetAlgoLined
\SetKwData{wbs}{wbs2}
\SetKwData{return}{return}

\SetKwProg{Fn}{Function}{:}{}

\KwIn{Data $\{X_t\}_{t = 1}^n$, number of intervals $R$, minimum interval length $I_{\text{min}}$, {threshold $D$, maximum scale $M$ for TAVC estimation}}

\BlankLine

\Fn{$\wbs(\{X_t\}_{t = 1}^n, s, e, R, D, M, I_{\text{min}}, \mc C)$}{
	\lIf{\textup{$e - s \leq I_{\text{min}}$}}{Quit}
	\BlankLine
	
	$\mc A_{s, e} \leftarrow \{(\ell, r) \in \Z^2:\, s \le \ell < r \le e \text{ and } r - \ell > 1\}$
	\BlankLine
	
	\uIf{\textup{$|\mc A_{s, e}| \leq R$}}{
	$\wt R \leftarrow |\mc A_{s, e}|$ and set	$\mc R_{s, e} \leftarrow \mc A_{s, e}$
	}
	\Else{$\wt R \leftarrow R$ and draw $\wt R$ elements from $\mc A_{s, e}$
	deterministically over an equispaced grid, to form $\mc R_{s, e} = \{(s_m,e_m) :\, 1 \le m \le \wt R \}$}
	\BlankLine
	
	\For{$m \in \{1, \ldots, \wt R\}$}{
    \uIf{\textup{$e_m - s_m \le M$}}{ %\vspace{3pt}
	Set $\wh{\sigma}^2_{s_m, e_m}$ as the solution to~\eqref{eq:m:est} with $L = 2 \lfloor (e_m - s_m)/2 \rfloor$
	} \Else{
	    Set $\wh{\sigma}^2_{s_m, e_m} \leftarrow \wh{\sigma}^2_M$
	    with $\wh{\sigma}^2_M$ solving~\eqref{eq:m:est} with $L = M$
	}
	\BlankLine
	
	Identify $(s_\circ, k_\circ, e_\circ) = \argmax_{(s_m, k, e_m): \, 1 \le m \le \wt R, \, s_m < k < e_m} \vert \mc T_{s_m, k, e_m} \vert / \wh{\sigma}_{s_m, e_m}$ 
	\BlankLine
	
	\lIf{\textup{$\vert \mc T_{s_\circ, k_\circ, e_\circ} \vert / \wh{\sigma}_{s_m, e_m} > D$}}{ Add $k_\circ$ to $\mc C$}
	\BlankLine
	
Perform $\wbs(\{X_t\}_{t = 1}^n, s, k_\circ,R, D, M, I_{\min}, \mc C) \cup \wbs(\{X_t\}_{t = 1}^n, k_\circ, e, R, D, M, I_{\min}, \mc C)$
	}
}
\BlankLine

Initialise $\mc C \leftarrow \emptyset$

$\wbs(\{X_t\}_{t = 1}^n, 0, n, R, D, M, I_{\min}, \mc C)$
\BlankLine

\KwOut{$\mc C$}
\end{algorithm}

\section{Additional numerical results}
\label{sec:add:sim}

In this section, we provide further information on the simulation study carried out in Section~\ref{sec:sim} and report additional numerical results to demonstrate the performance of the proposed robust estimator of the scale-dependent TAVC.

The covering metric (CM) used to assess the quality of the segmentation produced by the detected change point is defined as follows. The true change locations $\{\tau_i \}_{i=1}^{q}$ define a partition~$\mc P$ of the interval $\{1, 2 , \ldots , n \}$ into disjoint sets $\mc A_i$ such that $\mc A_i$ is the segment $\{ \tau_{i-1}+1, \ldots , \tau_i \}$. Similarly, the estimated change locations $\{\wh{\tau}_i \}_{i=1}^{\wh{q}}$ yield a partition $\wh{\mc P}$ of segments $\wh{\mc A}_i$. 
Then, CM is defined by
\begin{align*}
C(\wh{\mc P}, \mc P) = \frac{1}{n} \sum_{\mc A \in \mc P} |\mc A| \max_{\wh{\mc A} \in \wh{\mc P}} \left\{ \frac{|\mc A \cap \wh{\mc A}|}{| \mc A \cup \wh{\mc A}|}  \right\}.
\end{align*}
The CM takes values between $0$ and $1$, with a value of $1$ corresponding to a perfect segmentation, i.e.\ $\mc P = \wh{\mc P}$.

We repeat the simulations carried out in Section~\ref{sec:sim} with different values of $n \in \{500, 2000\}$. 
When $n = 500$, we introduce $q = 3$ change points to the time series at times $\tau_i = \lfloor (n/4) i \rfloor, \, i = 1, \ldots, 3$. For $n = 2000$, we have $q = 6$ change points at times $\tau_i = \lfloor (n/7) i \rfloor, \, i = 1, \ldots, 6$. Lastly, $\mu_i$ is set analogously as in the main text. 
See Tables~\ref{table:sim:two}--\ref{table:sim:three} for the results.

\begingroup
{\small
\setlength{\tabcolsep}{3pt}
\setlength{\LTcapwidth}{\textwidth}
\begin{longtable}{c c c ccccc cc}
\caption{Performance comparisons when $n = 500$. We report the size, the proportion of realisations where change points are falsely detected when $q = 0$,
and the distribution of the estimated number of change points, covering metric (CM) and relative MSE (RMSE) over 1000 realisations when $q = 3$.
The modal value of the distribution of the number of estimated change points for each method is shown in bold. The best performing method according to each metric when $q = 3$ is underlined.}
\label{table:sim:two}
\centering
\endfirsthead
\endhead
\toprule	 
&&& \multicolumn{5}{c}{$\wh{q} - q$} &   &   \\ 
 Model       & Method   & Size    &  $\leq-2$     & $-1$        & $\mathbf{0}$  & $1$    & $\geq 2$    & CM & RMSE       \\ 
\cmidrule(lr){1-2} \cmidrule(lr){3-3}  \cmidrule(lr){4-8} \cmidrule(lr){9-10}
        \ref{model-a}     &  MOSUM.TAVC$_{[1]}$   & 0.116  &  0.006     & 0.033   &  \bf{0.944}   & 0.017     & 0.000             & 0.944   & 7.188  \\
 &  MOSUM.TAVC$_{[2]}$  &  0.061    & 0.010    &  0.086   & \bf{0.897}  & 0.007      & 0.000        & 0.930 & 8.261 \\
 & WBS2.TAVC$_{[1]}$ & 0.068 & 0.008 & 0.041 & {\bf{0.951}} & 0.000 & 0.000 & {0.949} & {6.411}
  \\
  & WBS2.TAVC$_{[2]}$ & 0.037  & 0.036 & 0.092 & \bf{0.872} & 0.000 & 0.000 & 0.922 & 9.082  \\ \cmidrule(lr){3-3}  \cmidrule(lr){4-8} \cmidrule(lr){9-10}
       &  DepSMUCE(0.05) &  0.015   & 0.006   &  0.220    &  \bf{0.774} & 0.000      &  0.000         & 0.890  & 12.896 \\ 
     &  DepSMUCE(0.2) & 0.067    & 0.000   &  0.033    &  {\bf{0.965}} & 0.002      &  0.000         & 0.953  & 6.334 \\ 
     &  DeCAFS &  0.023  & 0.000   &  0.000    &  \bf{0.964} & 0.035      &  0.001         & {0.961}  & 6.062 \\ 
     &  WCM.gSa & 0.011    & 0.000   &  0.000    &  \bf{0.963} & 0.026     &  0.011         & 0.958  & {6.227} \\
     & MOSUM.oracle & 0.025 & 0.000 & 0.000 & \bf{0.879} & 0.115 &  0.006 & 0.948 & 6.863 \\
     & WBS2.oracle & 0.008 & 0.000 & 0.000 & \underline{\bf{0.997}} & 0.003 & 0.000 & \underline{0.963} & \underline{5.749} \\ \cmidrule(lr){1-2} \cmidrule(lr){3-3}  \cmidrule(lr){4-8} \cmidrule(lr){9-10}
     
         \ref{model-b}     &  MOSUM.TAVC$_{[1]}$  &  0.131  &  0.003     & 0.033   &  \bf{0.944}   & 0.020     & 0.000             & 0.946   & 7.020  \\
 &  MOSUM.TAVC$_{[2]}$ & 0.079      & 0.016    &  0.067   & \bf{0.906}  & 0.011      & 0.000 & 0.931 & 8.195 \\
 & WBS2.TAVC$_{[1]}$ & 0.077 & 0.017  & 0.033 & {\bf{0.950}} & 0.000 & 0.000 & {0.948}
 & {6.183} \\
  & WBS2.TAVC$_{[2]}$ & 0.040 & 0.042 &  0.081 & \bf{0.877} & 0.000 & 0.000 & 0.924 & 7.635
  \\ \cmidrule(lr){3-3}  \cmidrule(lr){4-8} \cmidrule(lr){9-10}
       &  DepSMUCE(0.05) & 0.371     & 0.003   &  0.153    &  \bf{0.643} & 0.125      &  0.076         & 0.883  & 20.031 \\ 
    &  DepSMUCE(0.2)&  0.572    & 0.000   &  0.022    &  \bf{0.609} & 0.200      &  0.169         & 0.917  & 15.960 \\ 
    &  DeCAFS & 0.746   & 0.000   &  0.000    &  0.232 & 0.096      &  \bf{0.672}         & 0.894  & 26.581 \\ 
     &  WCM.gSa &  0.013   & 0.000   &  0.000    &  {\bf{0.977}} & 0.022      & 0.001         &{0.959}  & {6.521} \\  
     & MOSUM.oracle & 0.040 & 0.000 & 0.000 & \bf{0.862} & 0.129 & 0.009 & 0.948 & 6.813 \\
     & WBS2.oracle & 0.015 & 0.000 & 0.000  & \underline{\bf{0.995}} & 0.004 & 0.001 &  \underline{0.964} & \underline{5.335} \\ \cmidrule(lr){1-2} \cmidrule(lr){3-3}  \cmidrule(lr){4-8} \cmidrule(lr){9-10}

\ref{model-c}     &  MOSUM.TAVC$_{[1]}$ &  0.168   &  0.000     & 0.003   &  \bf{0.992}   & 0.005     & 0.000             & 0.991   & 2.038  \\
 &  MOSUM.TAVC$_{[2]}$ & 0.113       & 0.000    &  0.016   & \bf{0.981}  & 0.003      & 0.000        & 0.989 & 2.153 \\
 & WBS2.TAVC$_{[1]}$ & 0.144 & 0.000 & 0.002 & {\bf{0.998}} & 0.000 & 0.000 & \underline{0.997} & 1.484 \\
  & WBS2.TAVC$_{[2]}$ & 0.096 & 0.001 & 0.017 & {\bf{0.982}} &0.000 & 0.000 & 0.992 & 1.698 
  \\ \cmidrule(lr){3-3}  \cmidrule(lr){4-8} \cmidrule(lr){9-10}
       &  DepSMUCE(0.05) & 0.963    & 0.000   &  0.000    &  {\bf{0.893}} & 0.106      &  0.001         & 0.987  & 1.841 \\ 
     &  DepSMUCE(0.2) & 0.993    & 0.000   &  0.000    &  \bf{0.750} & 0.218     &  0.032         & 0.972  & 2.196 \\ 
     &  DeCAFS &  0.574  & 0.000   &  0.002    &  \bf{0.646} & 0.276      &  0.078         & {0.976}  & \underline{1.302} \\ 
     &  WCM.gSa & 0.160    & 0.000   &  0.000    &  \bf{0.498} & 0.185      &  0.317         & 0.887  & 3.262 \\  
      & MOSUM.oracle & 0.001 & 0.000 & 0.001 & \bf{0.991} & 0.008 & 0.000 & 0.991 & 2.037 \\
     & WBS2.oracle & 0.000 & 0.000 & 0.000 & \underline{\bf{1.000}} & 0.000 & 0.000  & \underline{0.997} & 1.454 \\ \cmidrule(lr){1-2} \cmidrule(lr){3-3}  \cmidrule(lr){4-8} \cmidrule(lr){9-10}

\ref{model-d}     &  MOSUM.TAVC$_{[1]}$ & 0.130     &  0.001     & 0.016   & {\bf{0.977}}   & 0.006     & 0.000              & {0.979}   & 3.419  \\
 &  MOSUM.TAVC$_{[2]}$ & 0.071       & 0.005    &  0.051   & \bf{0.943}  & 0.001      & 0.000        & 0.968 & 4.715 \\
 & WBS2.TAVC$_{[1]}$ & 0.102 & 0.004 & 0.017 & {\bf{0.979}} & 0.000 & 0.000 & {0.986}
 & 2.091 \\
  & WBS2.TAVC$_{[2]}$ & 0.066 & 0.012 & 0.060 & \bf{0.928} & 0.000 & 0.000 & {0.971} & 2.905  \\ \cmidrule(lr){3-3}  \cmidrule(lr){4-8} \cmidrule(lr){9-10}
       &  DepSMUCE(0.05) & 0.762    & 0.000   &  0.000    &  \bf{0.959} & 0.039      &  0.002  & {0.988}  & 2.116 \\ 
     &  DepSMUCE(0.2) & 0.919     & 0.000   &  0.000    &  \bf{0.859} & 0.125      &  0.016         & 0.978  & 2.502 \\ 
    &  DeCAFS &  0.722  & 0.000   &  0.000    &  0.248 & 0.189      &  \bf{0.563}         & 0.903  & {2.642} \\ 
     &  WCM.gSa &  0.113   & 0.000   &  0.000    &  \bf{0.623} & 0.156      &  0.221         & 0.921  & 3.682 \\  
      & MOSUM.oracle & 0.003 & 0.000 & 0.001 & \bf{0.984} & 0.014 & 0.001 & 0.983 &  2.861 \\
     & WBS2.oracle & 0.001 & 0.000 & 0.000 & \underline{\bf{1.000}} & 0.000 & 0.000 & \underline{0.992} & \underline{1.776} \\ \cmidrule(lr){1-2} \cmidrule(lr){3-3}  \cmidrule(lr){4-8} \cmidrule(lr){9-10}

\ref{model-e}     &  MOSUM.TAVC$_{[1]}$ & 0.095    &  0.000    & 0.000   &  {\bf{0.965}}   & 0.035     & 0.000             & 0.985   &  78.044 \\
 &  MOSUM.TAVC$_{[2]}$ &  0.045      & 0.000    &  0.000   & {\bf{0.970}}  & 0.030      & 0.000        & 0.985 &  78.070 \\
 & WBS2.TAVC$_{[1]}$ & 0.191 & 0.000 & 0.000 & \underline{\bf{1.000}} & 0.000 & 0.000 & \underline{0.988} & \underline{66.844}  \\
  & WBS2.TAVC$_{[2]}$ & 0.098 & 0.000 & 0.000 & \underline{\bf{1.000}} & 0.000 & 0.000 & \underline{0.988} & \underline{66.844}
  \\ \cmidrule(lr){3-3}  \cmidrule(lr){4-8} \cmidrule(lr){9-10}
       &  DepSMUCE(0.05)&  0.968    & 0.000   &  0.000    &  0.237 & 0.125      &  \bf{0.638}         & 0.856  & 751.323 \\ 
     &  DepSMUCE(0.2) & 0.993     & 0.000   &  0.000    &  0.057 & 0.054      &  \bf{0.889}         & 0.754  & 1319.156 \\ 
     &  DeCAFS  & 0.004  & 0.000   &  0.000    &  \bf{0.983} & 0.014      &  0.003         & \underline{0.988}  & 84.649 \\ 
     &  WCM.gSa & 0.000    & 0.000   &  0.000    &  \underline{\bf{1.000}} & 0.000      &  0.000         & \underline{0.988}  & {67.557}  \\ 
 & MOSUM.oracle & 1.000 & 0.000 & 0.000 & 0.000 & 0.000 & \bf{1.000} & 0.284 & 250.212 \\
     & WBS2.oracle & 1.000 & 0.000 & 0.000 & 0.000 & 0.000 & \bf{1.000} & 0.289 & 201.271 \\ \cmidrule(lr){1-2} \cmidrule(lr){3-3}  \cmidrule(lr){4-8} \cmidrule(lr){9-10}

\ref{model-f}     &  MOSUM.TAVC$_{[1]}$ & 0.174   & 0.001       & 0.013  &  \bf{0.969}   & 0.016     & 0.001             & 0.958   & 7.016  \\
 &  MOSUM.TAVC$_{[2]}$ &  0.079     & 0.005    &  0.030   & {\bf{0.955}}  & 0.010      & 0.000        & 0.953 & 7.527 \\
 & WBS2.TAVC$_{[1]}$ & 0.105 & 0.007 & 0.017 & {\bf{0.975}} & 0.001 & 0.000 & {0.963} & {6.226} \\
  & WBS2.TAVC$_{[2]}$ & 0.058 & 0.015 & 0.048  & {\bf{0.937}} & 0.000 & 0.000 & 0.951 & {7.041}
  \\ \cmidrule(lr){3-3}  \cmidrule(lr){4-8} \cmidrule(lr){9-10}
       &  DepSMUCE(0.05) & 0.389     & 0.000   &  0.075    &  \bf{0.802} & 0.097     &  0.026        & 0.930  &  12.034 \\ 
     &  DepSMUCE(0.2) &  0.579    & 0.000   &  0.014    &  \bf{0.727} & 0.178      &  0.081        & 0.939  &  11.923 \\ 
     &  DeCAFS & 0.647   & 0.000   &  0.000   &  0.304 & 0.120      &  \bf{0.576}         & 0.906  & 24.573 \\ 
     &  WCM.gSa & 0.020    & 0.000   &  0.000    &  \bf{0.972} & 0.023      &  0.005         & {0.964}  & {6.510} \\ 
      & MOSUM.oracle & 0.025 & 0.000 & 0.000 & \bf{0.922} & 0.073 & 0.005 & 0.958 & 7.303 \\
     & WBS2.oracle & 0.006 & 0.000 & 0.000 & \underline{\bf{0.993}} & 0.006  & 0.001 & \underline{0.969} & \underline{5.939} \\ \cmidrule(lr){1-2} \cmidrule(lr){3-3}  \cmidrule(lr){4-8} \cmidrule(lr){9-10}

\ref{model-g}     &  MOSUM.TAVC$_{[1]}$ & 0.230   &  0.009     & 0.058  &  \bf{0.912}   & 0.021     & 0.000             & 0.940   & 6.810  \\
 &  MOSUM.TAVC$_{[2]}$ &  0.160     & 0.015    &  0.102   & {\bf{0.866}}  & 0.017      & 0.000        & 0.928 & 7.928 \\
 & WBS2.TAVC$_{[1]}$ & 0.205 & 0.017 & 0.087 & \bf{0.871} & 0.024 & 0.001 & {0.918} & {8.715} \\
  & WBS2.TAVC$_{[2]}$ & 0.141 & 0.038 & 0.124  & {\bf{0.823}} & 0.015 & 0.000 & 0.904 & {10.482}
  \\ \cmidrule(lr){3-3}  \cmidrule(lr){4-8} \cmidrule(lr){9-10}
       &  DepSMUCE(0.05) & 0.335     & 0.002   &  0.325    &  \bf{0.620} & 0.050     &  0.003        & 0.868  & 11.388  \\ 
     &  DepSMUCE(0.2) &  0.577    & 0.000   &  0.095    &  \bf{0.717} & 0.183      &  0.005        & 0.924  &  7.381 \\ 
     &  DeCAFS & 0.399   & 0.002   &  0.002   &  \bf{0.713} & 0.184      &  {0.099}         & 0.940  & 6.843 \\ 
     &  WCM.gSa & 0.284    & 0.054   &  0.095    &  \bf{0.741} & 0.074      &  0.036 & {0.908}  & {8.554} \\ 
      & MOSUM.oracle & 0.016 & 0.000 & 0.000 & \bf{0.924} & 0.073 & 0.003 & 0.954 & \underline{5.573} \\
     & WBS2.oracle & 0.006 & 0.000 & 0.000 & \underline{\bf{0.997}} & 0.003  & 0.000 & \underline{0.957} & {5.822} \\ \cmidrule(lr){1-2} \cmidrule(lr){3-3}  \cmidrule(lr){4-8} \cmidrule(lr){9-10}

\ref{model-h}     &  MOSUM.TAVC$_{[1]}$ & 0.247   & 0.007      & 0.058  &  \bf{0.897}   & 0.037     & 0.001             & 0.938   & 8.065  \\
 &  MOSUM.TAVC$_{[2]}$ &  0.164     & 0.013    &  0.087   & {\bf{0.877}}  & 0.021      & 0.002        & 0.929 & 8.329 \\
 & WBS2.TAVC$_{[1]}$ & 0.195 & 0.016 & 0.058 & \bf{0.898} & 0.028 & 0.000 & {0.922} & {9.555} \\
  & WBS2.TAVC$_{[2]}$ & 0.142 & 0.027 & 0.114  & {\bf{0.848}} & 0.011 & 0.000 & {0.906} & {10.293}
  \\ \cmidrule(lr){3-3}  \cmidrule(lr){4-8} \cmidrule(lr){9-10}
       &  DepSMUCE(0.05) & 0.180     & 0.063   &  \bf{0.712}    &  {0.213} & 0.012     &  0.000        & 0.748  & 16.963 \\ 
     &  DepSMUCE(0.2) &  0.397    & 0.009   &  0.434    &  \bf{0.482} & 0.072    &  0.003        & 0.832  & 12.487 \\ 
     &  DeCAFS & 0.301   & 0.006   &  0.023   &  \bf{0.662} & 0.199      &  {0.110}         & 0.930  &  9.065\\ 
     &  WCM.gSa & 0.236    & 0.040   &  0.232   &  \bf{0.692} & 0.022      &  0.014         & {0.885}  & {9.182} \\ 
      & MOSUM.oracle & 0.034 & 0.000 & 0.001 & \bf{0.863} & 0.131 & 0.005 & 0.947 & \underline{6.526} \\
     & WBS2.oracle & 0.014 & 0.000 & 0.000 & \underline{\bf{0.989}} & 0.011  & 0.000 & \underline{0.949} & {6.549} \\ \cmidrule(lr){1-2} \cmidrule(lr){3-3}  \cmidrule(lr){4-8} \cmidrule(lr){9-10}

\ref{model-i}     &  MOSUM.TAVC$_{[1]}$ & 0.261   & 0.005    &  0.251   & \bf{0.703}  & 0.040      & 0.001        & 0.886 & 8.747 \\
 &  MOSUM.TAVC$_{[2]}$ &  0.163     & 0.014    &  0.348   & \bf{0.615}  & 0.023      & 0.000        & 0.861 & 9.839 \\
 & WBS2.TAVC$_{[1]}$ & 0.178 & 0.010 & 0.329 & \bf{0.633} & 0.028 & 0.000 & 0.855 & 10.803 \\
  & WBS2.TAVC$_{[2]}$ & 0.117 & 0.025 & 0.440  & \bf{0.520} & 0.015 & 0.000 & 0.825 & 11.536
  \\ \cmidrule(lr){3-3}  \cmidrule(lr){4-8} \cmidrule(lr){9-10}
       &  DepSMUCE(0.05) & 0.121     & 0.058   &  \bf{0.924}    &  0.017 & 0.001     &  0.000        & 0.709  & 11.780 \\ 
     &  DepSMUCE(0.2) &  0.309    & 0.008   &  \bf{0.897}    &  0.084 & 0.011      &  0.000        & 0.725  & 11.289  \\ 
     &  DeCAFS & 0.084   & 0.004   & \bf{0.660}   &  0.224 & 0.081      &  0.031         & 0.774  & 11.858 \\ 
     &  WCM.gSa & 0.071    & 0.002   &  \bf{0.749}    &  0.219 & 0.019      &  0.011         & {0.775}  & {9.747} \\ 
      & MOSUM.oracle & 0.137 & 0.000 & 0.000 & {\bf{0.754}} & 0.216 & 0.030 & \underline{0.941} & \underline{7.328} \\
     & WBS2.oracle & 0.150& 0.000 & 0.000 & \underline{\bf{0.768}} & 0.228  & 0.004 & {0.917} & 8.366
\\
\bottomrule
\end{longtable}}
\endgroup

\begingroup
{\small
\setlength{\tabcolsep}{3pt}
\setlength{\LTcapwidth}{\textwidth}
\begin{longtable}{c c c ccccc cc}
\caption{Performance comparisons when $n = 2000$. We report the size, the proportion of realisations where change points are falsely detected when $q = 0$,
and the distribution of the estimated number of change points, covering metric (CM) and relative MSE (RMSE) over 1000 realisations when $q = 6$.
The modal value of the distribution of the number of estimated change points for each method is shown in bold. The best performing method according to each metric when $q = 6$ is underlined.}
\label{table:sim:three}
\endfirsthead
\endhead
\toprule	 
&&& \multicolumn{5}{c}{$\wh{q} - q$} &   &   \\ 
 Model       & Method   & Size    &  $\leq-2$     & $-1$        & $\mathbf{0}$  & $1$    & $\geq 2$    & CM & RMSE       \\ 
\cmidrule(lr){1-2} \cmidrule(lr){3-3}  \cmidrule(lr){4-8} \cmidrule(lr){9-10}
        \ref{model-a}     &  MOSUM.TAVC$_{[1]}$ & 0.126   &   0.000    & 0.000  &  \bf{0.983}   & 0.017     & 0.000             & 0.975   & 5.979  \\
 &  MOSUM.TAVC$_{[2]}$ &  0.068     & 0.000    &  0.005   & {\bf{0.991}}  & 0.004      & 0.000        & 0.974 & 6.185 \\
 & WBS2.TAVC$_{[1]}$ & 0.018 & 0.000 & 0.001 & {\bf{0.999}} & 0.000 & 0.000 & {0.981} & {4.781} \\
  & WBS2.TAVC$_{[2]}$ & 0.006 & 0.000 & 0.007  & {\bf{0.993}} & 0.000 & 0.000 & 0.981 & 4.851
  \\ 
\cmidrule(lr){3-3}  \cmidrule(lr){4-8} \cmidrule(lr){9-10}  
       &  DepSMUCE(0.05) & 0.007     & 0.000   &  0.000    &  \underline{\bf{1.000}} & 0.000     &  0.000        & \underline{0.982}  & \underline{4.764} \\ 
     &  DepSMUCE(0.2) &  0.067    & 0.000   &  0.000    &  \bf{0.998} & 0.001      &  0.001        & 0.981  & 4.769 \\ 
     &  DeCAFS & 0.006   &   0.000   &  0.000 & \bf{0.985}      &  0.014   & 0.001      & 0.981  & 4.844 \\ 
     &  WCM.gSa & 0.005    & 0.000   &  0.000    &  \bf{0.964} & 0.014      &  0.022         & {0.979}  & {6.087} \\ 
      & MOSUM.oracle & 0.040 & 0.000 & 0.000 & \bf{0.821} & 0.171 & 0.008 & 0.972 & 6.023  \\
     & WBS2.oracle & 0.000 & 0.000 & 0.000 & \underline{\bf{1.000}} & 0.000  & 0.000 & \underline{0.982} & {4.776} \\ \cmidrule(lr){1-2} \cmidrule(lr){3-3}  \cmidrule(lr){4-8} \cmidrule(lr){9-10}
     
     \ref{model-b}     &  MOSUM.TAVC$_{[1]}$ & 0.143   & 0.000      & 0.000  &  \bf{0.996}   & 0.004     & 0.000             & 0.975   & 5.940  \\
 &  MOSUM.TAVC$_{[2]}$ &  0.079     & 0.000    &  0.003   & {\bf{0.997}}  & 0.000      & 0.000       & 0.974 & 6.051 \\
 & WBS2.TAVC$_{[1]}$ & 0.025 & 0.000 & 0.002 & {\bf{0.998}} & 0.000 & 0.000 & \underline{0.983} & {4.432} \\
  & WBS2.TAVC$_{[2]}$ & 0.016 & 0.002 & 0.003  & {\bf{0.995}} & 0.000 & 0.000 & {0.982} & {4.459}
  \\ \cmidrule(lr){3-3}  \cmidrule(lr){4-8} \cmidrule(lr){9-10}
       &  DepSMUCE(0.05) & 0.828     & 0.000   &  0.000    &  {0.389} & 0.171     &  \bf{0.440}        & 0.941  & 16.822 \\ 
     &  DepSMUCE(0.2) &  0.924    & 0.000   &  0.000    &  0.209 & 0.142      &  \bf{0.649}        & 0.923  & 20.094 \\ 
     &  DeCAFS & 0.957   & 0.000   &  0.000   &  0.040 & 0.016      &  \bf{0.944}         & 0.879  & 31.810 \\ 
     &  WCM.gSa & 0.005    & 0.000   &  0.000    &  \bf{0.961} & 0.014      &  0.025         & {0.980}  & {4.686} \\ 
      & MOSUM.oracle & 0.051 & 0.000 & 0.000 & \bf{0.846} & 0.138 & 0.016 & 0.973 & 5.934 \\
     & WBS2.oracle & 0.002 & 0.000 & 0.000 & \underline{\bf{0.999}} & 0.001  & 0.000 & \underline{0.983} & \underline{4.413} \\ 
\cmidrule(lr){1-2} \cmidrule(lr){3-3}  \cmidrule(lr){4-8} \cmidrule(lr){9-10}
     
\ref{model-c}     &  MOSUM.TAVC$_{[1]}$ & 0.122   &  0.000     & 0.000  &  \bf{0.996}   & 0.004     & 0.000 & 0.995   & 2.050  \\
 &  MOSUM.TAVC$_{[2]}$ &  0.081     & 0.000    &  0.000   & {\bf{0.999}}  & 0.001      & 0.000 & 0.995 & 2.063 \\
 & WBS2.TAVC$_{[1]}$ & 0.034 & 0.000 & 0.000 & \underline{\bf{1.000}} & 0.000 & 0.000 & \underline{0.998} & {1.356} \\
  & WBS2.TAVC$_{[2]}$ & 0.017 & 0.000 & 0.003  & {\bf{0.997}} & 0.000 & 0.000 & \underline{0.998} & {1.394}
  \\ \cmidrule(lr){3-3}  \cmidrule(lr){4-8} \cmidrule(lr){9-10}
       &  DepSMUCE(0.05) & 0.925     & 0.000  & 0.000 &   \bf{0.975} & 0.025 &  0.000  & 0.997  &  1.410 \\ 
     &  DepSMUCE(0.2) &  0.991    & 0.000  &  0.000  &  \bf{0.887} & 0.109 &  0.004 & 0.992  &  1.575 \\ 
     &  DeCAFS & 0.591   & 0.000   &  0.000 &  \bf{0.605} & 0.345      & 0.050 & {0.988}  & \underline{1.176} \\ 
     &  WCM.gSa & 0.007    & 0.000 & 0.000 & \bf{0.643} & 0.150 &  0.207 & {0.959}  & {2.336} \\ 
      & MOSUM.oracle & 0.005 & 0.000 & 0.000 & \bf{0.974} & 0.026 & 0.000 & 0.994 & 1.980 \\
     & WBS2.oracle & 0.000 & 0.000 & 0.000 & \underline{\bf{1.000}} & 0.000 & 0.000 & \underline{0.998} & {1.356} \\ 
\cmidrule(lr){1-2} \cmidrule(lr){3-3}  \cmidrule(lr){4-8} \cmidrule(lr){9-10}

\ref{model-d}   &  MOSUM.TAVC$_{[1]}$ & 0.092   & 0.000      & 0.000  &  \bf{0.992}   & 0.008  & 0.000  & 0.990   & 3.017  \\
 &  MOSUM.TAVC$_{[2]}$ &  0.060     & 0.000    &  0.003   & {\bf{0.995}}  & 0.002      & 0.000        & 0.990 & 3.024 \\
 & WBS2.TAVC$_{[1]}$ & 0.020 & 0.000 & 0.001 & {\bf{0.999}} & 0.000 & 0.000 & \underline{0.996} & {1.808} \\
  & WBS2.TAVC$_{[2]}$ & 0.009 & 0.000 & 0.002  & {\bf{0.998}} & 0.000 & 0.000 & \underline{0.996} & 1.821
  \\ \cmidrule(lr){3-3}  \cmidrule(lr){4-8} \cmidrule(lr){9-10}
       &  DepSMUCE(0.05) & 0.559     & 0.000   &  0.000    &  \bf{0.999} & 0.001 &  0.000  & \underline{0.996}  & \underline{1.789} \\ 
     &  DepSMUCE(0.2) &  0.850    & 0.000   &  0.000    &  \bf{0.964} & 0.035  &  0.001 & 0.994  & 1.878 \\ 
     &  DeCAFS & 0.696   & 0.000   &  0.000   &  0.267 & 0.220      &  \bf{0.513}  & 0.961  & 1.944 \\ 
     &  WCM.gSa & 0.013    & 0.000   &  0.000    &  \bf{0.723} & 0.113  &  0.164 & {0.967}  & {2.803} \\ 
      & MOSUM.oracle & 0.004 & 0.000 & 0.000 & \bf{0.951} & 0.046 & 0.003 & 0.989 & 2.908 \\
     & WBS2.oracle & 0.000 & 0.000 & 0.000 & \underline{\bf{1.000}} & 0.000  & 0.000 & \underline{0.996} & {1.800} \\ 
     \cmidrule(lr){1-2} \cmidrule(lr){3-3}  \cmidrule(lr){4-8} \cmidrule(lr){9-10}

\ref{model-e}     &  MOSUM.TAVC$_{[1]}$ & 0.133   & 0.00      & 0.000  &  \underline{\bf{1.000}}   & 0.000     & 0.000  & 0.993   & 87.796  \\
 &  MOSUM.TAVC$_{[2]}$ &  0.071     & 0.000    &  0.000   & \underline{\bf{1.000}}  & 0.000      & 0.000 & 0.993 & 87.796 \\
 & WBS2.TAVC$_{[1]}$ & 0.066 & 0.000 & 0.000 & \underline{\bf{1.000}} & 0.000 & 0.000 & \underline{0.994} & \underline{74.530} \\
  & WBS2.TAVC$_{[2]}$ & 0.029 & 0.000 & 0.000  & \underline{\bf{1.000}} & 0.000 & 0.000 & \underline{0.994} & \underline{74.530}
  \\ \cmidrule(lr){3-3}  \cmidrule(lr){4-8} \cmidrule(lr){9-10}
       &  DepSMUCE(0.05) & 1.000 & 0.000  &  0.0000 & 0.000 & 0.000     &  \bf{1.000}  & 0.613  & 2496.320 \\ 
     &  DepSMUCE(0.2) &  1.000    & 0.000   &  0.000    &  {0.000} & 0.000      &  \bf{1.000} & 0.464  & 4116.486 \\ 
     &  DeCAFS & 0.000   & 0.000   &  0.000   &  \bf{0.993} & 0.007 & 0.000 & \underline{0.994}  & 77.484 \\ 
     &  WCM.gSa & 0.000    & 0.000   &  0.000  & \bf{1.000} & 0.000 &  0.000 & \underline{0.994}  & {74.648} \\ 
      & MOSUM.oracle & 1.000 & 0.000 & 0.000 & 0.000 & 0.000 & \bf{1.000} & 0.265 & 222.751 \\
     & WBS2.oracle & 0.978 & 0.000 & 0.000 & 0.000 & 0.000  & \bf{1.000} & 0.265 & 188.806 \\ \cmidrule(lr){1-2} \cmidrule(lr){3-3}  \cmidrule(lr){4-8} \cmidrule(lr){9-10}

\ref{model-f}  &  MOSUM.TAVC$_{[1]}$ & 0.192   &  0.000     & 0.002  &  \bf{0.988}   & 0.010     & 0.000 & 0.979   & 6.112  \\
 &  MOSUM.TAVC$_{[2]}$ &  0.117     & 0.000    &  0.002   & {\bf{0.995}}  & 0.003  & 0.000 & 0.979 & 6.259 \\
 & WBS2.TAVC$_{[1]}$ & 0.054 & 0.000 & 0.001 & \underline{\bf{0.999}} & 0.000 & 0.000 & \underline{0.985} & {4.641} \\
  & WBS2.TAVC$_{[2]}$ & 0.029 & 0.001 & 0.001  & {\bf{0.998}} & 0.000 & 0.000 & \underline{0.985} & {4.662}
  \\ \cmidrule(lr){3-3}  \cmidrule(lr){4-8} \cmidrule(lr){9-10}
       &  DepSMUCE(0.05) & 0.767     & 0.000   &  0.000    &  \bf{0.549} & 0.232 &  0.219  & 0.960  & 9.668 \\ 
     &  DepSMUCE(0.2) &  0.931    & 0.000   &  0.000    &  0.336 & 0.257&  \bf{0.407}   & 0.944  &  11.982 \\ 
     &  DeCAFS & 0.903   & 0.000   &  0.000   &  0.090 & 0.040   &  \bf{0.870}  & 0.911  & 28.388 \\ 
     &  WCM.gSa & 0.024    & 0.000   &  0.000 & \bf{0.958} & 0.017 &  0.025 & {0.983}  & {4.860} \\ 
      & MOSUM.oracle & 0.029 & 0.000 & 0.000 & \bf{0.892} & 0.101 & 0.007 & 0.978 & 6.151  \\
     & WBS2.oracle & 0.006 & 0.000 & 0.000 & {\bf{0.997}} & 0.002  & 0.001 & \underline{0.985} & \underline{4.631} \\ 
     \cmidrule(lr){1-2} \cmidrule(lr){3-3}  \cmidrule(lr){4-8} \cmidrule(lr){9-10}
     
     \ref{model-g}     &  MOSUM.TAVC$_{[1]}$ & 0.297   &  0.000     & 0.000  &  \bf{0.957}   & 0.043     & 0.000    & 0.979   & 4.854 \\
 &  MOSUM.TAVC$_{[2]}$ &  0.214     & 0.000    &  0.001   & {\bf{0.971}}  & 0.027      & 0.001        & 0.979 & 4.840 \\
 & WBS2.TAVC$_{[1]}$ & 0.178 & 0.000 & 0.000 & \bf{0.992} & 0.008 & 0.000 & {0.979} & {4.959} \\
  & WBS2.TAVC$_{[2]}$ & 0.120 & 0.000 & 0.001  & \underline{\bf{0.997}} & 0.002 & 0.000 & 0.979 & {4.984}
  \\ \cmidrule(lr){3-3}  \cmidrule(lr){4-8} \cmidrule(lr){9-10}
       &  DepSMUCE(0.05) & 0.445     & 0.004   &  0.292    &  \bf{0.616} & 0.088 &  0.000   & 0.917  & 9.351 \\ 
     &  DepSMUCE(0.2) &  0.726    & 0.000  &  0.035    &  \bf{0.670} & 0.284 &  0.011   & 0.963  & 5.326 \\ 
     &  DeCAFS & 0.557   & 0.000   &  0.000   &  \bf{0.492} & 0.216 &  0.292  & 0.961  & 6.242  \\ 
     &  WCM.gSa & 0.316    & 0.031   &  0.027    &  \bf{0.895} & 0.026 &  0.021 & {0.968}  & {4.840} \\ 
      & MOSUM.oracle & 0.036 & 0.000 & 0.000 & \bf{0.878} & 0.111 & 0.011 & 0.975 & 4.933 \\
     & WBS2.oracle & 0.005 & 0.000 & 0.000 & \underline{\bf{0.997}} & 0.003  & 0.000 & \underline{0.983} & \underline{4.154} \\ 
     \cmidrule(lr){1-2} \cmidrule(lr){3-3}  \cmidrule(lr){4-8} \cmidrule(lr){9-10}

\ref{model-h}    &  MOSUM.TAVC$_{[1]}$ & 0.304   & 0.000   & 0.000  &  \bf{0.939}   & 0.058     & 0.003 & 0.976   & 5.985 \\
 &  MOSUM.TAVC$_{[2]}$ &  0.195     & 0.000    &  0.000   & {\bf{0.960}}  & 0.039   & 0.001  & 0.976 & 5.934 \\
 & WBS2.TAVC$_{[1]}$ & 0.142 & 0.000 & 0.000 & \bf{0.994} & 0.006 & 0.000 & 0.976 & {5.958} \\
  & WBS2.TAVC$_{[2]}$ & 0.081 & 0.000 & 0.001  & \underline{\bf{0.998}} & 0.001 & 0.000 & 0.977 & 5.796
  \\ \cmidrule(lr){3-3}  \cmidrule(lr){4-8} \cmidrule(lr){9-10}
       &  DepSMUCE(0.05) & 0.277     & 0.077   &  \bf{0.704}    &  0.210 & 0.009     &  0.000   & 0.833  & 15.628  \\ 
     &  DepSMUCE(0.2) &  0.595    & 0.004   &  0.361    &  \bf{0.514} & 0.113  &  0.008   & 0.901  & 10.853 \\ 
     &  DeCAFS & 0.368   & 0.000   &  0.000   &  \bf{0.594} & 0.199 &  0.207 & 0.965  & 7.261  \\ 
     &  WCM.gSa & 0.136    & 0.040   &  0.051    &  \bf{0.894} & 0.010 &  0.005 & {0.962}  & {5.811} \\ 
      & MOSUM.oracle & 0.060 & 0.000 & 0.000 & \bf{0.805} & 0.174 & 0.021 & 0.971 & 6.190 \\
     & WBS2.oracle & 0.004 & 0.000 & 0.000 & \bf{0.997} & 0.003  & 0.000 & \underline{0.980} & \underline{5.162} \\ 
     \cmidrule(lr){1-2} \cmidrule(lr){3-3}  \cmidrule(lr){4-8} \cmidrule(lr){9-10}

\ref{model-i}     &  MOSUM.TAVC$_{[1]}$ & 0.382   &  0.000     & 0.003  &  \bf{0.922}   & 0.072     & 0.003 & \underline{0.974}   & 6.551  \\
 &  MOSUM.TAVC$_{[2]}$ &  0.259     & 0.000    &  0.004   & {\bf{0.953}}  & 0.042 & 0.001   & \underline{0.974} & \underline{6.549} \\
 & WBS2.TAVC$_{[1]}$ & 0.231 & 0.000 & 0.005 & \bf{0.985} & 0.010 & 0.000 & 0.973 & 7.309 \\
  & WBS2.TAVC$_{[2]}$ & 0.181 & 0.000 & 0.003  & \underline{\bf{0.993}} & 0.004 & 0.000 & \underline{0.974} & {7.153}
  \\ \cmidrule(lr){3-3}  \cmidrule(lr){4-8} \cmidrule(lr){9-10}
       &  DepSMUCE(0.05) & 0.231     & \bf{0.741}   &  0.251    & 0.008 & 0.000 &  0.000 & 0.731  & 13.329 \\ 
     &  DepSMUCE(0.2) &  0.540    & 0.269   &  \bf{0.616}    &  0.106 & 0.008 &  0.001 & 0.784  & 13.165 \\ 
     &  DeCAFS & 0.081   & 0.262   &  0.022   &  \bf{0.574} & 0.092      &  0.050  & 0.890  & 8.834 \\ 
     &  WCM.gSa & 0.061    & 0.281   &  0.076    &  \bf{0.624} & 0.014  &  0.005 & {0.890}  & {8.122} \\ 
      & MOSUM.oracle & 0.097 & 0.000 & 0.000 & \bf{0.754} & 0.215 & 0.031 & 0.968 & 6.883  \\
     & WBS2.oracle & 0.067 & 0.000 & 0.000 & {\bf{0.957}} & 0.042  & 0.001 & 0.967 & 6.774
     \\
\bottomrule
\end{longtable}}
\endgroup

\section{Proof of Theorem~\ref{thm:one}}
\label{sec:proof}

For sequences of positive numbers $\{a_n\}$ and $\{b_n\}$, we write $a_n \lesssim b_n$, or $a_n = \mc O (b_n)$, if there exists some constant $C > 0$
such that $a_n/b_n \le C$ as $n \to \infty$. We write $a_n \asymp b_n$ if there exists some positive constants $C_1$ and $C_2$ such that $C_1 \leq a_n/b_n \leq C_2$ as $n \to \infty$. 
Without loss of generality, we set $b = 0$ and drop the dependence on $b$ for notational simplicity;
analogous arguments are applicable when other fixed values of $b \in \{0, \ldots, G - 1\}$ are used.

We denote by $\mc B_j = \{ t \in \{ 1, \ldots, n \} : \, jG+1 \leq t \leq (j+1)G \}$ the set of indices of the $j$-th block of data for some $j = 0, \ldots, \lfloor n/G \rfloor - 1$.
We adapt the proof of Theorem~5 in \cite{chen2021inference} with modifications to our case with the TAVC estimator. We denote by
\begin{align*}
\mc S = \{0 \le j \le \lfloor n/G \rfloor - 1 : \, \mc B_j \text{ or } \mc B_{j-1} \text{ contains change points} \}.   \end{align*}
Then, $\vert \mc S \vert \leq 2q$. 
First, we consider the influence function constructed using the blocks which do not contain any change points:
\begin{align*}
\bar{h}_L(u)  = \frac{1}{ N_0} \sum_{j \notin \mc S} \phi_v( \xi_j - u ) = 0, \text{ where } N_0 = N_1 - |\mc S|.
\end{align*}
Letting
\begin{align*}
\tilde{\xi} = \frac{1}{N_0} \sum_{j \notin \mc S} \mathbb{E} (\xi_j ) = { \wt{\sigma}_L^2}, \text{ and } \gamma^2 = \frac{1}{N_0} \sum_{j \notin \mc S} \mathbb{E}(\xi_j^2) - \tilde{\xi}^2, 
\end{align*}
define the functions 
\begin{align*}
B^{+} (u,x) &= \tilde{\xi} - u + \frac{v}{2} \left[ \left(\tilde{\xi}-u \right)^2 + \gamma^2 \right] +x,    \\
B^{-} (u,x) &= \tilde{\xi} - u - \frac{v}{2} \left[ \left(\tilde{\xi}-u \right)^2 + \gamma^2 \right] -x.
\end{align*}
Then, it can be shown that 
$\mathbb{E}(\bar{h}_L (u) )$ satisfies the envelope property
\begin{align}
\label{eq:no-change-envelope}
  B^{-} (u,0)    \leq  \mathbb{E} \left(\bar{h}_L (u) \right)   \leq B^{+} (u,0).   
\end{align}
To see this, note that since $\phi(x) \leq x +x^2/2$ by Equation~\eqref{eq:envelope}, we have 
\begin{align*}
\mathbb{E}\left( \bar{h}_L (u) \right) & \leq \frac{1}{v N_0} \sum_{j \notin \mc S} \l [ \mathbb{E} ( v ({\xi}_j -u)) + \frac{1}{2}\mathbb{E} ( v^2 ({\xi}_j -u)^2) \r ] \\
& = \frac{1}{N_0} \sum_{j \notin \mc S} \mathbb{E} ( \xi_j) - u + \frac{v}{2N_0} \sum_{j \notin \mc S} \mathbb{E} ( (\xi_j - u)^2) \\
& = \tilde{\xi} -u + \frac{v}{2}\l[ (\tilde{\xi} -u)^2 + \gamma^2 \r] = B^{+} (u,0).
\end{align*}
Similarly, from the fact that $\phi(x) \geq x - x^2/2$, the lower bound follows.

Next, we show that $\bar{h}_L (u)$ is concentrated about its mean. We deal with the different cases, Assumptions~\ref{assum:noise}~\ref{assum:noise:three}~\ref{cond:moment} and~\ref{cond:subexp}, separately, in order to prove Equations~\eqref{eq:thm:one} and \eqref{eq:thm:two} respectively.
Applying Step~2 of the proof of Theorem~5 in \cite{chen2021inference}, 
we have that, for $C_0 > 0$ and $x > 0$, 
\begin{align}
\label{eq:delta-net}
\p\left(  \sup_{u: \, |u-\wt{\sigma}_L^2 | \leq C_0} \left| \bar{h}_L(u) -\mathbb{E}(\bar{h}_L(u))  \right| \geq x /N_0   \right) \leq \p\left(  \max_{u \in A_n} \left| \bar{h}_L(u) -\mathbb{E}(\bar{h}_L(u))  \right| \geq x / (2N_0)   \right),
\end{align}
where $A_n$ is the $\delta$-net for $\{ u: | u - \wt{\sigma}^2_L | \leq C_0 \}$ with $\delta = \wt{\sigma}^2_L x/ (2N_0)$ and $|A_n| = \mathcal{O}(N_0/x)$. 
%{\cred In what follows, we write $\sup_u$ in place of $\sup_{u: \, |u-\wt{\sigma}_L^2 | \leq C_0}$.} 
For any random variable $X$, denote by $\mathbb{E}_0 (X) = X- \mathbb{E}(X)$ the centering operator, and let {$\Delta \bar{X}_j = \bar{X}_j - \bar{X}_{j-1}$}. Let $\mc F_k = \{\eta_t, \,  t \in \cup_{l \leq k} \mc B_l \}$, and {$\mc F_{k, \{ m \}}$}, $m \leq k$,
%\footnote{initially $\mc F_{s, \{ \ell \}}$} 
be $\mc F_k$ with $\eta_t$ therein replaced with its independent and identically distributed copy $\eta^\prime_t$ for all $t \in \mc B_m$. 
Then for any random variable $X = h (\mc F_k)$ with measurable $h(\cdot)$, 
let $X_{\{m\}} = h(\mc F_{k, \{m\}})$. 

\begin{proof}[Proof of~\eqref{eq:thm:one}]
Under Assumption~\ref{assum:noise}~\ref{assum:noise:three}~\ref{cond:moment},
we show that $\zeta_j (u) = \phi_{v} (\xi_j -u)$ satisfies
appropriate functional dependence properties.
Denote
\begin{align*}
\delta_{\ell,r/2} = \max_{j \notin \mc S} \l\Vert  \sup_{u} \vert \zeta_j (u) - \zeta_{j, \{j-\ell\}} (u) \vert \r\Vert_{r/2}.
\end{align*}
%We must show that $\delta_{\ell,r/2}$ decays at a suitable rate, such that under Assumption~\ref{assum:noise}~\ref{assum:noise:three}~\ref{cond:moment}, we have that
%\begin{align*}
%w_{\alpha,r/2} = \sup_{k\geq 0} \ (k+1)^{\alpha} \sum_{\ell=k}^\infty \delta_{\ell,r/2} < \infty.    
%\end{align*}
%Showing this result will allow us to apply Lemma C.2 in the supplementary material of \cite{zhang2017gaussian}. To this end, note that 
Since $ |\phi' |_{\infty} \leq 1$, we have for any $\ell \in \mathbb{N}$ and $u \in \mathbb{R}$,
\begin{align}
& \l\Vert \ \sup_u \l\vert \zeta_j (u) - \zeta_{j,\{j-\ell\}} (u) \r\vert \r\Vert_{r/2}   \leq \l\Vert \xi_j - \xi_{j,\{j-\ell\} } \r\Vert_{r/2} = \frac{G}{2} \Vert \Delta \bar{X}_j^2 -  \Delta \bar{X}_{j,\{ j-\ell\}}^2  \Vert_{r/2} \nonumber \\
& \leq \frac{G}{2} \l(  \l\Vert \mathbb{E}_0 \left[ \Delta \bar{X}_{j} ( \Delta \bar{X}_j - \Delta \bar{X}_{j,\{j-\ell\}} ) \right] \r\Vert_{r/2} + 
\l\Vert \mathbb{E}_0 \left[ \Delta \bar{X}_{j, \{j-\ell\}} ( \Delta \bar{X}_j - \Delta \bar{X}_{j,\{j-\ell\}} ) \right] \r\Vert_{r/2} \r) \nonumber \\ 
& = \frac{G}{2} \left( I_1 + I_2 \right).  
\label{eq:func:dep}
\end{align}
Further, let $E_{j,k} = \sum_{t=jG+1}^{(j+1)G} a_{t-k}$
and $U_j = G^{-1} \sum_{t=jG+1}^{(j+1)G} \varepsilon_t$. Then,
\begin{align*}
U_j = \frac{1}{G} \sum_{t=jG+1}^{(j+1)G} \sum_{k=0}^\infty a_{k} \eta_{t-k} = \frac{1}{G} \sum_{t=jG+1}^{(j+1)G} \sum_{k=-\infty}^t a_{t-k} \eta_k = \frac{1}{G}\sum_{k \leq (j+1)G} E_{j,k} \eta_{k},
\end{align*}
and we define
\begin{align*}
\Delta U_j =  \frac{1}{G} \sum_{k \leq (j+1)G} E_{j,k} \eta_{k} - \frac{1}{G} \sum_{k \leq jG} E_{j-1, k} \eta_{k}.   
\end{align*}
Noting that 
\begin{align*}
G \Delta U_{j,\{j-\ell \}} =  
\l\{\begin{array}{ll}
\displaystyle{\sum_{\substack{k \leq (j+1)G, \\ k \notin \mc B_{j -\ell}}} E_{j,k} \eta_k + \sum_{k \in \mc B_{j-\ell}} E_{j,k} \eta_k' - \sum_{\substack{k \leq jG, \\ k \notin \mc B_{j -\ell}}} E_{j-1,k} \eta_k - \sum_{k \in \mc B_{j-\ell}} E_{j-1,k} \eta_k'}  & \text{ for } \ell \geq 1, \\
\displaystyle{\sum_{k \leq jG} E_{j,k} \eta_k + \sum_{k \in \mc B_j} E_{j,k} \eta_k' - \sum_{k \leq jG} E_{j-1,k} \eta_k} & \text{ for } \ell =0,
\end{array}\r.
\end{align*}
it follows that
\begin{align*}
G \l(\Delta U_j - \Delta U_{j, \{ j- \ell \}} \r) =
\l\{\begin{array}{ll}
\displaystyle{\sum_{k \in \mc B_{j-\ell}}} ( E_{j,k} - E_{j-1,k} ) ( \eta_k - \eta'_k  )  & \text{ for } \ell \geq 1, \\
\displaystyle{\sum_{k \in \mc B_{j}}} E_{j,k}  ( \eta_k - \eta'_k  )  & \text{ for } \ell =0 .
\end{array}\r.
\end{align*}
Next, by Assumption~\ref{assum:noise}~\ref{assum:noise:one},
we have
\begin{align}
\label{eq:decay}
\l\vert \sum_{k \ge \ell} a_k \r\vert
\le \sum_{k \ge \ell} \vert a_k \vert
\le \Xi \sum_{k \ge \ell} (1 + k)^{-\beta}
\lesssim (1 + \ell)^{- \beta + 1 + \epsilon}
\end{align}
for any arbitrarily small $\epsilon > 0$.
Then, using~\eqref{eq:decay},
\begin{align}
\sum_{k \in \mc B_{j-\ell}} E_{j-1,k}^2 &= \sum_{k = (j-\ell)G+1 }^{(j-\ell+1)G} \left( \sum_{t=(j-1)G+1}^{jG} a_{t-k} \right)^2 \nonumber \\
& \lesssim \sum_{k = (j-\ell)G+1 }^{(j-\ell+1)G} \left( (j-1)G + 1-k  \right)^{-2(\beta - 1 - \epsilon)} \lesssim G(G(\ell-1) \vee 1)^{-2(\beta - 1 - \epsilon)},
\label{eq:bound:e:sq:one} \\
\sum_{k \leq (j+1)G } E_{j,k}^2 &= \sum_{k \leq (j+1)G } \left( \sum_{t=jG+1}^{(j+1)G} a_{t-k} \right)^2 \lesssim \sum_{k \leq (j+1)G } \left(1 \vee (jG+1-k) \right)^{-2 (\beta - 1 - \epsilon)} \lesssim G,
\label{eq:bound:e:sq:two}
\end{align}
where the constants involved in $\lesssim$ depend on $\epsilon$, $\beta$ and $\Xi$. Since we only consider $j \notin \mc S$, 
we bound $I_1$ in~\eqref{eq:func:dep} for $\ell \geq 1$ as
\begin{align*}
I_1 &=  \left\Vert \mathbb{E}_0 \left[ \Delta U_{j} ( \Delta U_j - \Delta U_{j,\{j-\ell\}} ) \right] \right\Vert_{r/2} \\
&= G^{-2} \left\Vert \mathbb{E}_0 \left[ \left( \sum_{k \leq (j+1)G} E_{j,k} \eta_k - \sum_{k \leq jG} E_{j-1,k} \eta_k \right) \left( \sum_{k \in \mc B_{j-\ell}} ( E_{j,k} - E_{j-1,k} ) ( \eta_k - \eta'_k  ) \right) \right] \right\Vert_{r/2} \\
& \lesssim G^{-2} \left\Vert \sum_{k \leq (j+1)G} E_{j,k} \eta_k - \sum_{k \leq jG} E_{j-1,k} \eta_k \right\Vert_{r} \left\Vert \sum_{k \in \mc B_{j-\ell}} ( E_{j,k} - E_{j-1,k} ) ( \eta_k - \eta'_k ) \right\Vert_{r} \\
& \lesssim G^{-2} C_r^2 \left( \sum_{k \leq (j+1)G} E_{j,k}^2 \mu_r^2 + \sum_{k \leq jG} E_{j-1,k}^2 \mu_r^2 \right)^{1/2} \left( \sum_{k \in \mc B_{j-\ell}}  ( E_{j,k} - E_{j-1,k} )^2 \mu_r^2  \right)^{1/2} \\
& \lesssim G^{-1} C_r^2 \left( G( \ell-1) \vee 1 \right)^{-(\beta - 1 - \epsilon)} \mu_r^2 ,
\end{align*}
where the first inequality follows from H\"{o}lder's inequality,
the second from Burkholder's inequality (see e.g.\ Lemma~2 of \cite{chen2021inference}) with $C_r = \max(\sqrt{r - 1}, 1/(r - 1))$,
the last from~\eqref{eq:bound:e:sq:one}--\eqref{eq:bound:e:sq:two}
and $\mu_r = \Vert \eta_1 \Vert_r$. 
Similarly, when $\ell =0$, we have that $I_1 \lesssim G^{-1} C_r^2 \mu_{r}^2$. We can bound the term $I_2$ in~\eqref{eq:func:dep} in a similar fashion, to obtain
\begin{align}
\label{eq:delta-s}
\delta_{\ell,r/2} \lesssim \left( (G\ell)^{-(\beta - 1 - \epsilon)} \mathbb{I}(\ell > 1) + \mathbb{I}(\ell\leq 1) \right) C_r^2 \mu_r^2.
\end{align}
Therefore, the dependence-adjusted norm
\begin{align*}
w_{r/2, \alpha} = \sup_{k\geq 0} \ (k+1)^{\alpha} \sum_{\ell=k}^\infty \delta_{\ell,r/2} \lesssim  \sup_{k\geq 2} \ (k+1)^{\alpha} \sum_{\ell=k}^\infty (G\ell)^{-(\beta - 1 - \epsilon)} C_r^2 \mu_r^2 \lesssim C_r^2 \mu_r^2,
\end{align*}
for $\alpha = \beta - 2 - 2\epsilon > 1/2-2/r$ with $\beta > 2.5$. 
Under Assumption~\ref{assum:noise}~\ref{assum:noise:three}~\ref{cond:moment}, having shown the bound on $w_{r/2, \alpha}$, 
we apply Lemma~C.2 of \cite{zhang2017gaussian}
and yield, for any $u \in A_n$, 
\begin{align*}
\p \left(\left| \bar{h}_L(u) -\mathbb{E}(\bar{h}_L(u))  \right| \geq \frac{x}{2N_0}   \right) \lesssim  N_0 x^{-r/2} + \exp\l(-\frac{x^2}{cN_0} \r),
\end{align*}
where $c$ and the constant appearing in the $\lesssim$ 
depend on $r$, $\mu_r$, $\Xi$, $\beta$ and $\epsilon$.
Applying a Bonferroni bound with $|A_n| = \mathcal{O}(N_0/x)$ then yields 
\begin{align}
\label{eq:concentration:one}
\p \left(  \max_{u \in A_n} \left| \bar{h}_L(u) -\mathbb{E}(\bar{h}_L(u))  \right| \geq \frac{x}{2N_0} \right) \lesssim \frac{N_0}{x} \left( N_0 x^{-r/2} + \exp\l(-\frac{x^2}{cN_0} \r) \right).
\end{align}
Taking $x \asymp N_0^{\frac{4}{r + 2}} \vee ( N_0 \log(N_0))^{1/2}$, 
%\footnote{you meant $r$ not $v$ in $N_0^{\frac{4}{r + 2}}$?}
using Equations~\eqref{eq:delta-net} and \eqref{eq:concentration:one} we obtain 
\begin{align}\label{eq:op:one}
     \sup_{|u-\wt{\sigma}_L^2 | \leq C_0} \left| \bar{h}_L(u) -\mathbb{E}(\bar{h}_L(u))  \right| = \mc O_P \left( N_0^{-\frac{r - 2}{r + 2}} \vee \sqrt{\frac{\log(N_0)}{N_0} } \right).
\end{align}
Recalling that $| \mc S| \leq 2q$ and $| \phi(x) | \leq \log(2)$, we have that
\begin{align}
\label{eq:bound:one}
\left| \frac{N_1 h_L (u)}{N_0} - \bar{h}_L (u) \right| \leq \frac{2q \log(2)}{v N_0}.    
\end{align}
Combining Equations~\eqref{eq:no-change-envelope}, \eqref{eq:op:one} and~\eqref{eq:bound:one}, we have that with probability tending to one,
\begin{align}
\label{eq:bound:two}
B^- (u, \Delta) \leq \frac{N_1 h_L (u)}{N_0} \leq B^+ (u, \Delta) 
\end{align}
uniformly for all $|u - \wt{\sigma}^2_L| \leq C_0$,
where
\begin{align*}
\Delta = \frac{x}{N_0} + \frac{2q \log(2)}{v N_0},
\end{align*}
with $x \asymp N_0^{\frac{4}{r + 2}} \vee (N_0 \log(N_0))^{1/2}$.
If
\begin{align}
\label{eq:prop:one}
v^2 \gamma^2 + 2v \Delta \leq 1, 
\end{align}
then $B^+ (u, \Delta)$ possesses real roots. Denote the smallest root by $u^+$, which satisfies 
\begin{align}
\label{property2}
u^+ \leq \tilde{\xi} + \frac{\gamma^2 + 2\Delta/v}{\sqrt{1/v^2}}
\le \tilde{\xi} + v\gamma^2 + 2 \Delta.
\end{align}
Under Assumption~\ref{assum:noise}, we have for some $j \notin \mc S$,
\begin{align*}
\gamma^2 &\le \E(\xi_j^2) = \frac{G^2}{4} \E(\Delta \bar{X}_j^4)
= \frac{G^2}{4} \E(\Delta U_j^4)
\lesssim \frac{1}{G^2} \l(\l\Vert \sum_{k \le (j + 1)G} E_{j, k} \eta_k \r\Vert_4^4
+ \l\Vert \sum_{k \le jG} E_{j - 1, k} \eta_k \r\Vert_4^4\r)
\\
&\lesssim \frac{1}{G^2} \l( \sum_{k \le (j + 1)G} E_{j, k}^2 + \sum_{k \le jG} E_{j - 1, k}^2 \r)^2 \mu_4^4 = \mathcal{O}(1)
\end{align*}
by~\eqref{eq:bound:e:sq:two} and Burkholder's inequality.
Therefore, from~\eqref{property2} and that $u^+ \ge \wt\xi = \wt{\sigma}_L^2$,
\begin{align*}
u^+ - \wt{\sigma}_L^2 = \mc O \left( v + \frac{x}{N_0} + \frac{q}{v N_0} \right) = \mc O \left( v + \frac{Gx}{n} + \frac{Gq}{vn} \right).  
\end{align*}
A similar bound can be obtained for $u^-$, the largest root of $B^{-1}(u, \Delta)$
and under Equation~\eqref{eq:bound:two}, we have that $u^- \leq \wh{\sigma}^2_L \leq u^+$. Then, setting $v \asymp (Gq/n)^{1/2}$, we have~\eqref{eq:prop:one} holds, and
\begin{align*}
\l\vert \wh{\sigma}^2_L - \wt{\sigma}_L^2 \r\vert = \mathcal{O}_P\l( \sqrt{ \frac{Gq}{n} } + \max\l\{ \l( \frac{G}{n} \r)^{\frac{r-2}{r+2}}, \sqrt{\frac{G \log(n)}{n}} \r\} \r).
\end{align*}
\end{proof}

\begin{proof}[Proof of~\eqref{eq:thm:two}]
We proceed analogously as in the proof of~\eqref{eq:thm:one}, 
except that we control for the dependence-adjusted sub-exponential norm
of $\zeta_j(u)$, as
%\begin{align*}
%\tilde{w}_f = \sup_{r \geq 2} r^{-f} \sum_{\ell=0}^{\infty} \delta_{\ell,r}
%\end{align*}
% in order to apply Lemma C.4 of \cite{zhang2017gaussian}. 
\begin{align*}
\tilde{w}_f = \sup_{r \geq 2} r^{-f} \sum_{\ell=0}^{\infty} \delta_{\ell,r} \lesssim \ \sup_{r \geq 2} r^{-f} \sum_{\ell=0}^{\infty}  \left( (G\ell)^{-(\beta - 1 - \epsilon)} \mathbb{I}(\ell > 1) + \mathbb{I}(\ell \leq 1) \right) C_{2r}^2 \mu_{2r}^2
\lesssim \sup_{r \geq 2} r^{-f + 2\kappa + 1},
\end{align*}
where the first inequality follows from~\eqref{eq:delta-s}
and the second from Assumption~\ref{assum:noise}~\ref{assum:noise:one}.
Therefore, we have $\tilde{w}_f < \infty$ with $f = 2\kappa + 1$. 
Then applying Lemma~C.4 of \cite{zhang2017gaussian} 
with~\eqref{eq:delta-net}, we obtain
\begin{align*}
\p \left(  \sup_{|u-\wt{\sigma}_L^2 | \leq C_0} \left| \bar{h}_L(u) -\mathbb{E}(\bar{h}_L(u))  \right| \geq \frac{x}{N_0}   \right) \lesssim \frac{N_0}{x} \exp \left( -\frac{(x/\sqrt{N_0})^{\frac{2}{4\kappa + 3}}}{c} \right),
\end{align*}
where $c$ depends on $\kappa$ and $\beta$ through $\tilde{w}_f$.
Taking $x \asymp \log^{2\kappa + 3/2}(n) N_0^{1/2}$, we have that
\begin{align}
\label{eq:op:two}
\sup_{|u-\wt{\sigma}_L^2 | \leq C_0} \left| \bar{h}_L(u) -\mathbb{E}(\bar{h}_L(u))  \right| = \mc O_P \left( \frac{\log^{2\kappa + 3/2} (n)}{\sqrt{N_0}} \right).
\end{align}
Then, by the analogous arguments as those adopted in the proof of~\eqref{eq:thm:one} with~\eqref{eq:op:two} replacing~\eqref{eq:op:one}, we derive~\eqref{eq:thm:two}.
\end{proof}

\begin{proof}[Proof of~\eqref{eq:thm:three}]
%First, we show that 
%\begin{align}\label{approx-bias}
%|\wt{\sigma}^2_L - \sigma_L^2| = \mathcal{O} \left( \min \left\{ \max \{ L^{-1}, L^{1-\beta} \}, L^{-\beta/(\beta+1)} \right\} \right) .  
%\end{align}
Recalling that $U_j = G^{-1} \sum_{t = jG + 1}^{(j + 1)G} \vep_t$, we have
\begin{align*}
\wt{\sigma}^2_L = \frac{G}{2} \mathbb{E}\l( (U_j - U_{j-1})^2 \r) &= \frac{G}{2} \mathbb{E}(U_j^2 + U_{j-1}^2 - 2U_j U_{j-1})
= \sigma^2_{G} - G \mathbb{E}(U_{j} U_{j-1}),
\end{align*}
while
\begin{align*}
{\sigma}^2_L &=  \mathbb{E}\left( \left( \frac{1}{\sqrt{L}} \sum_{t=1}^{L} \varepsilon_t \right)^2\right) = \mathbb{E} \left( \left(  \frac{1}{\sqrt{L}} \sum_{t=1}^{G} \varepsilon_t + \frac{1}{\sqrt{L}}\sum_{t=G+1}^{L} \varepsilon_t \right)^2   \right) 
\\
&= \frac{1}{2} (\sigma^2_G + \sigma^2_G) + \frac{2}{L} \mathbb{E} \left( \sum_{t=1}^{G} \varepsilon_t  \sum_{t'=G+1}^{L} \varepsilon_{t'}\right)
= \sigma^2_G + G  \mathbb{E} (U_j U_{j-1}).
\end{align*}
By Assumption~\ref{assum:noise}~\ref{assum:noise:one},
$\rho_k = \mathbb{E} (\varepsilon_0 \varepsilon_k)$, $k \ge 0$, satisfy
\begin{align*}
| \rho_k | = \l\vert \sum_{j=0} a_j a_{j+k} \r\vert \leq 
\sum_{j=0}^\infty \vert a_j a_{j + k} \vert
\lesssim \sum_{j=0}^\infty (j + 1)^{-\beta} (j + k + 1)^{-\beta}
\le k^{-\beta} \sum_{j=0}^\infty (j + 1)^{-\beta}
= \mathcal{O}(k^{-\beta}).
\end{align*}
Therefore, 
\begin{align*}
|\wt{\sigma}^2_L - \sigma_L^2  | &= L | \mathbb{E} (U_j U_{j-1}) |
= \frac{4}{L} \sum_{t = 1}^G \sum_{t' = G + 1}^{L} \E(\vep_t \vep_{t'})
= \frac{4}{L} \left( \sum_{k=1}^{G} k \rho_k + \sum_{k=1}^{G-1}(G- k) \rho_{k+G}  \right)
\\
&\lesssim L^{-1} \left( \sum_{k=1}^{G} k^{1-\beta} + \sum_{k=1}^{G-1}(G- k)(k+G)^{-\beta}   \right)
\lesssim L^{-1} \left( 1 + L^{2-\beta} \right) 
 = \mathcal{O}\left( L^{-1} \right).
\end{align*}
Lastly, to prove the second statement in Equation~\eqref{sigma-approx2}, 
we show that
\begin{align*}
| \sigma^2_L - \sigma^2 | \le \l\vert \sum_{k: \, \vert k \vert \geq L} \rho_k \r\vert
+ L^{-1} \l\vert \sum_{k = -L+1}^{L-1} \vert k \vert \rho_k \r\vert
= \mathcal{O}(L^{-\beta + 1} + L^{-1}) = \mathcal{O}(L^{-1}),
\end{align*}
which completes the proof. 
\end{proof}

\end{document}